\documentclass[a4paper,twocolumn,11pt,accepted=2024-02-02]{quantumarticle}
\pdfoutput=1
\usepackage[utf8]{inputenc}
\usepackage[english]{babel}
\usepackage[T1]{fontenc}
\usepackage{xr}
\usepackage{amsmath}
\usepackage{amsfonts,amssymb,amsthm,amsxtra,dsfont}
\usepackage{physics}
\usepackage[numbers,sort&compress]{natbib}
\usepackage[caption=false]{subfig}
\usepackage[]{graphicx}
\usepackage{grffile}
\usepackage{mathrsfs}
\usepackage{xcolor}
\usepackage{mathtools}
\usepackage[ruled]{algorithm2e}
\usepackage{tikz}
\usetikzlibrary{quantikz}
\usepackage[bookmarks=false,colorlinks=true,urlcolor=blue,citecolor=blue,linkcolor=blue]{hyperref}
\usepackage{diagbox}
\usepackage{chngcntr}

\usepackage{makecell}
\usepackage{xcolor}
\usepackage{soul}
\makeatletter
\DeclareFontFamily{OMX}{MnSymbolE}{}
\DeclareSymbolFont{MnLargeSymbols}{OMX}{MnSymbolE}{m}{n}
\SetSymbolFont{MnLargeSymbols}{bold}{OMX}{MnSymbolE}{b}{n}
\DeclareFontShape{OMX}{MnSymbolE}{m}{n}{
    <-6>  MnSymbolE5
   <6-7>  MnSymbolE6
   <7-8>  MnSymbolE7
   <8-9>  MnSymbolE8
   <9-10> MnSymbolE9
  <10-12> MnSymbolE10
  <12->   MnSymbolE12
}{}
\DeclareFontShape{OMX}{MnSymbolE}{b}{n}{
    <-6>  MnSymbolE-Bold5
   <6-7>  MnSymbolE-Bold6
   <7-8>  MnSymbolE-Bold7
   <8-9>  MnSymbolE-Bold8
   <9-10> MnSymbolE-Bold9
  <10-12> MnSymbolE-Bold10
  <12->   MnSymbolE-Bold12
}{}

\let\llangle\@undefined
\let\rrangle\@undefined
\DeclareMathDelimiter{\llangle}{\mathopen}%
                     {MnLargeSymbols}{'164}{MnLargeSymbols}{'164}
\DeclareMathDelimiter{\rrangle}{\mathclose}%
                     {MnLargeSymbols}{'171}{MnLargeSymbols}{'171}
\makeatother
\newcommand{\kett}[1]{\lvert #1 \rrangle}
\newcommand{\Heff}{H_\mathrm{eff}}
\newcommand{\He}{H_\mathrm{e}}
\newcommand{\Ha}{H_\mathrm{a}}
\newcommand{\bM}{\mathbf{M}}
\newcommand{\bV}{\mathbf{V}}
\newcommand{\bth}{\boldsymbol{\theta}}
\newcommand{\Lag}{\mathcal{L}}
\newcommand{\be}{\begin{equation}}
\newcommand{\ee}{\end{equation}}
\SetKwComment{Comment}{/* }{ */}

\begin{document}

\title{Adaptive variational simulation for open quantum system dynamics}

\author{Huo Chen}
\email{huochen@lbl.gov}
\orcid{0000-0002-8405-4596}
\affiliation{Computational Research Division, Lawrence Berkeley National Laboratory, Berkeley, California 94720, USA}

\author{Niladri Gomes}
\email{niladri@lbl.gov}
\orcid{0000-0003-2762-6866}
\thanks{\\ H.C. and N.G. contributed equally.}
\affiliation{Computational Research Division, Lawrence Berkeley National Laboratory, Berkeley, California 94720, USA}

\author{Siyuan Niu}
\email{siyuanniu@lbl.gov}
\orcid{0000-0003-4683-381X}
\author{Wibe Albert de Jong}
\email{WAdeJong@lbl.gov}
\orcid{0000-0002-7114-8315}
\affiliation{Computational Research Division, Lawrence Berkeley National Laboratory, Berkeley, California 94720, USA}

\maketitle

\begin{abstract}
Emerging quantum hardware provides new possibilities for quantum simulation. While much of the research has focused on simulating closed quantum systems, the real-world quantum systems are mostly open. Therefore, it is essential to develop quantum algorithms that can effectively simulate open quantum systems.
Here we present an adaptive variational quantum algorithm for simulating open quantum system dynamics described by the Lindblad equation. The algorithm is designed to build resource-efficient ans\"atze through the dynamical addition of operators by maintaining the simulation accuracy. We validate the effectiveness of our algorithm on both noiseless simulators and IBM Q quantum processors and observe good quantitative and qualitative agreement with the exact solution. 
We also investigate the scaling of the required resources with system size and accuracy and find polynomial behavior. 
%We also provide numerical evidence that the resource required for our algorithm does not scale exponential with respect to either the system size or accuracy for small systems. 
Our results demonstrate that near-future quantum processors are capable of simulating open quantum systems.
\end{abstract}

\section{Introduction}
The theory of open quantum system investigates the behavior of small quantum systems in contact with a large environment~\cite{Breuer2002-yt, Weiss2012-vk, lidar_lecture_notes}. Recently there has been a surge of interest in developing efficient simulation algorithms for open quantum systems~\cite{Weimer2021-ys, Benjamin_prl_2020, Hu2020-cd, Wang2023-sa, Suri2023twounitary}. On one hand, the emergence of quantum technologies for building artificial quantum systems for computing~\cite{De_Leon2021-bd, Nielsen2002-jq}, sensing~\cite{Degen2017-wu,Marciniak2022-cq}, light harvesting~\cite{Collini2010-hs,Mattioni2021-av}, and other applications promises an abundant scientific and societal benefits. These artificial quantum systems must be treated and designed as open because perfect isolation of a quantum system is extremely challenging, if not impossible.  
On the other hand, the theory of open quantum systems plays a crucial role in the fundamental science because it describes many non-equilibrium processes, including quarkonium suppression in heavy-ion collisions~\cite{Yao2021-nx}, relaxation dynamics of many-body quantum system~\cite{Weimer2021-ys}, charge and energy transfer dynamics in molecular systems~\cite{May2011-ka}, and others. All of these applications will benefit from improved simulation algorithms for open quantum systems.

Similar to the case of closed system, the complexity of classical algorithms for simulating open system scales exponentially with the system size. As the quantum computers hold the promise of being able to efficiently simulate quantum systems, it is worth exploring the new opportunities offered by the rapid advancement of cloud-accessible quantum computers~\cite{PhysRevA.94.032329, De_Jong2021-ap, Metcalf2020-jo, Maslov2021-xz, Bassman2021-sh, Urbanek2021-vz, Klymko2022-zs, Harper2019-yb, Pokharel2023-aj, bibek_2022_grover} for open-system simulations. Current research mostly is focused on solving the 
Gorini-Kossakowski-Sudarshan-Lindblad (GKSL)  equation (or simply Lindblad equation) due to its broad applications in various subfields of quantum systems engineering~\cite{Kossakowski1972-vj,Lindblad1976-qn,Gorini1978-aq,Breuer2002-yt,lidar_lecture_notes}. 
Based on the underlying techniques, existing quantum algorithms for simulating the Lindblad dynamics fall into five categories: unitary dilation~\cite{Hu2020-cd,Hu2022-su,Rost2021-co,De_Jong2021-ap}, variational simulation~\cite{Benjamin_prl_2020}, quantum imaginary-time evolution (QITE)~\cite{Kamakari2022-yr}, Monte Carlo~\cite{jose_2023} and qubit-bath engineering~\cite{juha_2022,Metcalf2020-jo,Wang2011-wg}. Among them the variational simulation and QITE require less circuit depth and are considered more NISQ (noisy intermediate-scale quantum)~\cite{Preskill2018-xb} friendly.

Variational algorithms have been shown to be capable of solving real and imaginary time evolutions \cite{shen2017, mcardle2019variational, AVQITE}, excited states \cite{zhang2021adaptive}, thermal states \cite{getelina2023}, non-Hermitian quantum mechanics for non-equilibrium systems \cite{fogedby1995continuum}, open quantum systems and  general first-order differential equations~\cite{Benjamin_prl_2020}. In this paper, we present a compact approach for solving the Lindblad equation using a time-dependent adaptive variational method. Our strategy to generate NISQ-friendly ans\"atze is built upon adaptive variational quantum dynamics simulations (AVQDS)~\cite{Yao2021-is} and its imaginary-time counterpart (AVQITE) \cite{AVQITE}. Distinct from previous works, our focus is on the dynamics of open quantum systems, rather than the closed system dynamics or ground state solutions typically explored. The main idea is to reformulate the Lindblad equation as a Schr\"odinger equation with an effective non-Hermitian Hamiltonian, and utilizing a quantum computer to simulate the evolution of the normalized state vector while employing a classical computer to record its norm. The presence of both unitary and dissipative elements in the equation complicates the adaptive process, as it lacks a pre-defined lower bound, rendering the establishment of a practical threshold impractical. To address this challenge, we have developed an unrestricted adaptive protocol, specifically tailored to circumvent the difficulties in setting a feasible threshold for the McLachlan distance.
% I am hesitant to put it this way because we did seperate the Hamiltonian and Lindblad propagator through Trotterization. It could cause more confusion.
%In simple terms, because the system dynamics involve both a loss of energy (dissipation) meaning a mix of unitary and non-unitary evolution, we can use the AVQDS method to simulate the unitary part and apply a concept similar to AVQITE to simulate the non-unitary part.
%A physical interpretation of the non-unitary evolution is illustrated in Fig.~\ref{fig:bloch_sphere}, which we believe also provides insight into how AVQITE works. 
%\textbf{BERT: You should point out the other novel innovative advancements that were needed here} 

To showcase the efficacy of our algorithm, we simulate the open quantum system dynamics of a quantum annealing (QA) process, focusing on the alternating-sector-chain (ASC) problem as described in~\cite{Mishra2018-ff, Reichardt2004-kd}.
QA is a quantum meta-heuristic algorithm that is commonly implemented on analog quantum hardware. Although our algorithm is specifically designed for gate-based quantum computers, simulating QA hardware highlights its applicability to a wider range of use cases, such as analyzing artificial quantum systems. Furthermore, the ASC problem stands out as an effective toy problem to benchmark open-system algorithms. On one hand, the system is integrable in the closed-system case because its Hamiltonian can be transformed into a free-fermion model by Jordan-Wigner transformation. On the other hand, the introduction of local dephasing and amplitude damping noise perturb the free-fermion model and make it a non-trivial task to simulate the quantum dynamics. Given that dephasing and amplitude damping represent two of the most prevalent quantum noise models, holding significant relevance in real-world applications~\cite{Hu2022-su, Kamakari2022-yr, jose_2023}, the open-system ASC problem serves as a ideal test platform for open-system algorithms.

We demonstrate our algorithm on both the noiseless simulator and IBM quantum hardware. Our results show good quantitative and  qualitative agreement with the exact solution. Furthermore, we perform numerical analysis of the resource requirement to simulate small systems, and find polynomial scaling with respect to either the system size or the desired accuracy. These results provide compelling evidence that open quantum system simulation is within the capability of near-term quantum devices.
 
%\begin{itemize}
%\color{red}
%    \item Both simulation and real hardware results show good agreement with the exact solution.
%    \item The resourced needed does not scale exponentially.
%\end{itemize}

In summary, we propose a new adaptive variational quantum algorithm tailored for simulating the Lindblad master equation, where prior adaptive approaches encounter the challenge of establishing a feasible threshold for the McLachlan distance due to the coexistence of unitary and dissipative components. We validate the effectiveness of our algorithm on both simulators and IBM quantum hardware. Our results show that open quantum system simulation is within the capability of near-term quantum devices. The structure of this paper is as follows. In Sec.~\ref{sec:method}, we introduce the Lindblad equation and describe our quantum algorithm,  unrestricted adaptive variational
quantum dynamics (UAVQDS), for solving it. In Sec.~\ref{sec:noiseless_dim} we report the performance and resource requirement of our algorithm based on the results from the noiseless simulator.
In Sec.~\ref{sec:hardware}, we report the performance of our algorithm on IBM quantum computers. We conclude in Sec.~\ref{sec:conclusion}, and present additional technical details and derivations in the Appendices.

\section{Method}
\label{sec:method}
For ease of reference, a table summarizing frequently used symbols can be found in Appendix~\ref{app:symbols}.
\subsection{The Lindblad master equation}
When a quantum system interacts with its environment, we can use the Lindblad master equation (ME) to model its behavior. Unlike closed quantum systems where a single quantum state (pure state) can describe the system's complete information, open quantum systems require a density matrix, which is a statistical ensemble of multiple quantum states (mixed state) to fully describe the system's behavior. Here, we focus on the Lindblad equation in the diagonal form
\begin{equation}
    \label{eq:Lindblad_eq}
    \frac{d}{dt}\rho\pqty{t} = -i\comm{H\pqty{t}}{\rho\pqty{t}}+\Lag\bqty{\rho\pqty{t}} \ ,
\end{equation}
where $\rho\pqty{t}$ is the density matrix and $H$ is the Hamiltonian describing the system of interest. The environment density matrix is traced out while deriving the Lindblad equation and its interaction with the system (in the weak coupling regime) is taken care of by a dissipative term $\Lag\bqty{\rho\pqty{t}}$ which is given by the form,
\begin{equation}
    \Lag\bqty{\rho\pqty{t}} = \sum_{k=1}^{K} \gamma_k \pqty{L_k \rho\pqty{t} L^\dagger_k -\frac{1}{2}\acomm{L^\dagger_k L_k}{\rho\pqty{t}}} \ .
\end{equation}
The operators $L_k$ are known as the Lindblad operators, and are taken to be dimensionless. $\gamma_k$ are the dissipation rates, which are non-negative and possess a dimension of inverse time. %Since the Lindblad equation is a dissipative model, the rate of dissipation or decay of the system over time is modeled by varying $\gamma_k$. The $\gamma_k$ are non-negative quantities with dimension of inverse time. 
Usually it is convenient to absorb $\gamma_k$ into the definition of $L_k$,
\begin{equation}
    \sqrt{\gamma_k} L_k \mapsto L_k \ .
\end{equation}
Throughout our discussions, we have adopted the convention of setting $\hbar = 1$.

Because a quantum computer only evolves a pure state under unitaries, the density matrix equation (Eq.~\eqref{eq:Lindblad_eq}) cannot be directly solved by a gate-based quantum computer.   This challenge can be addressed by two different approaches: 1) vectorizing~\cite{horn1991topics} or 2) stochastic unravelling of the Lindblad ME~\cite{Breuer2002-yt,Yip2018-dc, Brun2002-bt,Gardiner2004-qa}. We briefly review here how these two methods work.

\textit{Vectorization} -- The core idea of vectorization is to convert the density matrix to a vector by stacking the columns of $\rho$ on top of one another:
\begin{equation}
    \mathrm{vec}\pqty{\rho} \equiv \bqty{\rho_{11}, \rho_{12}\cdots\rho_{21},\rho_{22}\cdots\rho_{\mathrm{D}\mathrm{D}}}^\mathrm{T} \ ,
\end{equation}
where $\mathrm{D}\times \mathrm{D}$ is the dimension of $\rho$. We will also use the symbol $\rvert \rho \rrangle$ interchangeably with $\mathrm{vec}\pqty{\rho}$. The map $\mathrm{vec}$ is a linear isometry between the $\mathrm{D}\times \mathrm{D}$ Liouville space (Hilbert space under the Hilbert-Schmidt inner product) of $\rho$ and the $\mathrm{D}^2$ Hilbert space of $\rvert \rho \rrangle$, with the preserved norms being the trace norm and $L_2$ norm
\begin{equation}
    \label{eq:iso_trace}
    \sqrt{\Tr\pqty{\rho^\dagger\rho}} = \sqrt{\llangle \rho \vert\rho \rrangle} \ .
\end{equation}

Using the identity $\operatorname{vec}(A B C)=\left(C^{\mathrm{T}} \otimes A\right) \operatorname{vec}(B)$, Eq.~\eqref{eq:Lindblad_eq} can be rewritten as
\begin{equation}
    \frac{d}{dt}\kett{\rho\pqty{t}} = -iH_{\mathrm{vec}}\pqty{t} \kett{\rho\pqty{t}} \ ,
\end{equation}
where $H_{\mathrm{vec}}$ is an effective non-Hermitian Hamiltonian given by (see Appendix~\ref{appendix:vectorized_lindblad} for the derivation)
\begin{align}
    \label{eq:Heff_vec}
    &H_\mathrm{vec} = I\otimes H - H^T\otimes I \notag\\
    &+i\sum_k \bqty{\bar{L}_k\otimes L_k-\frac{1}{2}\pqty{I\otimes L_k^{\dagger} L_k + L_k^{T} \bar{L}_k\otimes I}} \ ,
\end{align}
where $\bar{L}_k$ denotes the conjugate of $L_k$ and $I$ is the identity matrix matching the dimension of $H$.
We will refer to this approach as the vectorization method for the remainder of the paper.

The vectorization method is a well-established technique for the classical simulation of open quantum systems. However, its application within quantum algorithms remains largely underexplored. To the best of our knowledge, reference~\cite{Kamakari2022-yr} stands as the sole work discussing this methodology in the context of simulating open system dynamics on quantum computers. In this cited work, the authors employ QITE to solve the vectorization Lindblad equation and chose ans\"atze based on numerical experiments. Contrarily, our study presents a systematic approach to construct resource-efficient ans\"atze, enabling us to execute larger simulations on IBM quantum computers.

An alternative way to vectorize the ME is to expressing the density matrix in the basis of Pauli matrices and representing the superoperators through the Pauli transfer matrices (PTMs). However, a straightforward application of this technique could significantly increase the circuit complexity, making it less suitable for demonstrations on NISQ hardware. We provide an discussion of this approach in Appendix~\ref{appendix:vectorization_pauli_basis}.

\textit{Stochastic unravelling} --
The Lindblad master equation can be unravelled using quantum trajectory method~\cite{Breuer2002-yt,Yip2018-dc, Brun2002-bt,Gardiner2004-qa}. For each trajectory, the evolution is governed by a deterministic evolution and a jump process. The deterministic evolution is described by the Schr\"odinger equation associated with a non-Hermitian Hamiltonian
\begin{equation}
    \label{eq:unravelled_equation}
    \frac{d}{dt}\ket{\tilde{\psi}\pqty{t}} = -i H_\mathrm{urv}\ket{\tilde{\psi}\pqty{t}} \ ,
\end{equation}
where $\ket{\tilde{\psi}\pqty{t}}$ is the unnormalized state vector and the effective Hamiltonian is given by
\begin{equation}
    H_\mathrm{urv} = H\pqty{t} -\frac{i}{2}\sum_{k=1}^{K} L^\dagger_k L_k \ .
\end{equation}

For an infinitesimal time step from $t$ to $t+dt$, there are two possible evolutions for $\ket{\tilde{\psi}\pqty{t}}$: either the deterministic evolution subject to Eq.~\eqref{eq:unravelled_equation} (with probability $1-dp$) or a jump occurring (with probability $dp$). The probability of a jump is given by
\begin{equation}
    dp = \sum_{k=1}^K \bra{\tilde{\psi}\pqty{t}} L_k^\dagger L_k \ket{\tilde{\psi}\pqty{t}} / \braket{\tilde{\psi}\pqty{t}} dt \ .
\end{equation}

Furthermore, if a jump happens, the unnormalized state is updated as
\begin{equation}
    \ket{\tilde{\psi}\pqty{t+dt}} = L_i\ket{\tilde{\psi}\pqty{t}}/\sqrt{\mel{\tilde{\psi}\pqty{t}}{L^\dagger_i L_i}{\tilde{\psi}\pqty{t}}}
\end{equation}
where $L_i$ is randomly picked from $\Bqty{L_k}_{k=1}^K$ with probability
\begin{equation}
    \label{eq:jump_p}
    p_i =     \mel{\tilde{\psi}\pqty{t}}{L^\dagger_i L_i}{\tilde{\psi}\pqty{t}}/\sum_{k=1}^K\mel{\tilde{\psi}\pqty{t}}{L^\dagger_k L_k}{\tilde{\psi}\pqty{t}} \ .
\end{equation}

Let us denote the normalized state vector by $\ket{\psi_j\pqty{t}}=\ket{\tilde{\psi}_j\pqty{t}} / \sqrt{\braket{\tilde{\psi}_j\pqty{t}}{\tilde{\psi}_j\pqty{t}}}$ where the index $j$ represent the $j$th trajectory. The density matrix solution to Eq.~\eqref{eq:Lindblad_eq} can be constructed by
$\rho\pqty{t}=\frac{1}{n}\sum_{j=1}^n \dyad{\psi_j\pqty{t}}{\psi_j\pqty{t}}$ for large enough $n$. We will refer to this approach as the trajectory method henceforth.

\subsection{Solving the effective Schr\"odinger equation}
The key step in both the aforementioned methods is to solve an effective Schr\"odinger equation with a non-Hermitian Hamiltonian, which can be accomplished using variational quantum algorithms~\cite{Yuan2019-ei, endo2020}. 
Without loss of generality, we use $H_\mathrm{eff}$ to represent the effective Hamiltonian in both methods, and $\ket{\tilde{\psi}\pqty{t}}$, $\ket{\psi\pqty{t}}$ to represent the unnormalized and normalized state vector solutions. Furthermore, we split $H_\mathrm{eff}$ into its Hermitian and anti-Hermitian parts and denotes them by $H_\mathrm{e}$ and $H_\mathrm{a}$ respectively
\be
\label{eq:split_H}
H_\mathrm{eff} = \frac{H_\mathrm{eff}+H^\dagger_\mathrm{eff}}{2} -i\pqty{ i\frac{H_\mathrm{eff}-H^\dagger_\mathrm{eff}}{2}} \equiv H_\mathrm{e} - i H_\mathrm{a} \ .
\ee
Inspired by AVQDS~\cite{Yao2021-is}, we propose a similar adaptive protocol for the open quantum system simulation. We briefly describe the non-adaptive algorithm here and leave the adaptive protocol to next section. The core idea of an variational algorithm is to encode quantum states with a sequence of parameterized circuits
\begin{equation}
    \label{eq:ansatz}
    \ket{\phi\pqty{t}} = \prod_{\mu=1}^{k} e^{-i\theta_\mu\pqty{t} A_\mu} \ket{\psi_\mathrm{R}} \ ,
\end{equation}
where $e^{-i\theta_\mu\pqty{t} A_\mu}$ is the $\mu$th layer of circuit controlled by the real %\textbf{BERT:Does it have to be real?} 
parameter $\theta_\mu\pqty{t}$, and $\ket{\psi_\mathrm{R}}$ is a fixed reference state. Then the evolution of the state can be mapped to
the evolution of the parameters controlling the circuit, i.e., $\theta_\mu\pqty{t}$ in Eq.~\eqref{eq:ansatz}. For adaptive protocols, the ansatz operators $A_\mu$ are adaptively added from an ansatz pool at runtime.

Our algorithm is based on McLachlan variational
principle~\cite{MacDonald1934-mv}
\be
\label{eq:var_prin}
\delta \norm{\frac{d\ket{\phi\pqty{\bth\pqty{t}}}}{dt}+iH_\mathrm{eff}\ket{\phi\pqty{\bth\pqty{t}}}}^{2} =0 \ ,
\ee
where $\bth\pqty{t}$ denotes $\bqty{\theta_1\pqty{t}, \theta_2\pqty{t}, \cdots \theta_k\pqty{t}}$. The norm in Eq.~\eqref{eq:var_prin} is known as the McLachlan distance and we denote its square by $\mathcal{D}$. The corresponding evolution equation of the variational parameters is given by
\be
\label{eq:theta_eom}
\bM \dot{\bth} = \bV \ ,
\ee
where $\bM$ is a matrix with elements
\be
\label{eq:M}
    \bM_{\mu \nu} =2\mathrm{Re}\pqty{\pdv{\bra{\phi}}{\theta_{\mu}}\pdv{\ket{\phi}}{\theta_{\nu}} +\bra{\phi}\pdv{\ket{\phi}}{\theta_{\mu}} \bra{\phi}\pdv{\ket{\phi}}{\theta_{\nu}}} \ ,
\ee
and 
$\bV$ is a vector with elements
\be
\label{eq:V}
    \bV_\mu = 2\mathrm{Im}\pqty{\expval{\Heff}\bra{\phi}\pdv{\ket{\phi}}{\theta_\mu}+\pdv{\bra{\phi}}{\theta_\mu}\Heff \ket{\phi}} \ ,
\ee
where $\expval{\Heff} = \mel{\phi}{\Heff}{\phi}$. Detailed derivation of the evolution equation can be found in the Appendix~\ref{appendix:variational_alg}. Assuming every ansatz operator is a single Pauli string, each term in Eqs.~\eqref{eq:M} and~\eqref{eq:V} involving the derivatives can be evaluated using a quantum computer by either direct or indirect measurements~\cite{Mitarai2019-pv,Yao2021-is} (See Appendix~\ref{app:MV_measurement} for a description of the relevant quantum circuits). 

\begin{figure}
    \centering
    \includegraphics[width=0.7\linewidth]{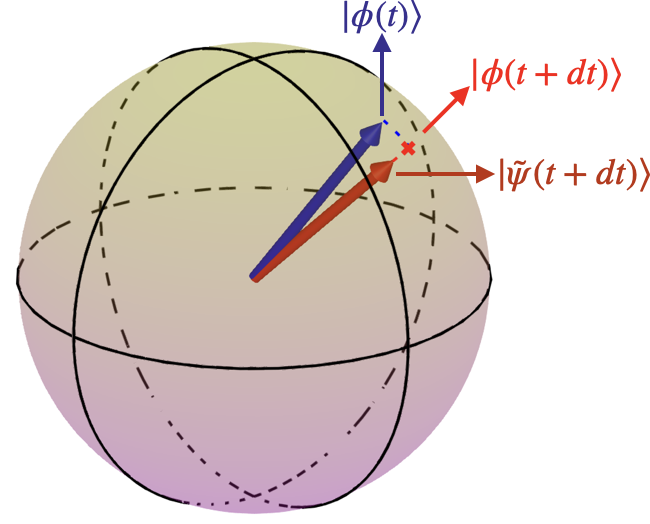}
    \caption{\textbf{An illustration of how a non-unitary evolution is simulated using an unitary.} Assume at time $t$ the state is a pure state $\ket{\phi\pqty{t}}$ on the Bloch sphere (blue arrow). After an infinitesimal step $dt$, an evolution subject to a non-Hermitian effective Hamiltonian will bring the state inside the Bloch sphere (red arrow) by shrinking its norm $\braket{\tilde{\psi}}{\tilde{\psi}} < 1$. The goal of the variational algorithm is to evolve $\ket{\phi\pqty{t}}$ to the closest state to $\ket{\tilde{\psi}\pqty{t+dt}}$ on the Bloch sphere (red cross), %i.e., $\ket{\phi\pqty{t+dt}}= \ket{\tilde{\psi}\pqty{t+dt}} / \norm{\ket{\tilde{\psi}\pqty{t+dt}}}$,
    i.e., $\ket{\phi\pqty{t+dt}}= \ket{\psi\pqty{t+dt}} $,
    while recording $\norm{\ket{\tilde{\psi}\pqty{t+dt}}}$ as classical information.}
    \label{fig:bloch_sphere}
\end{figure}

Before we proceed, it is important to note that our algorithm effectively utilizes only the ansatz state (Eq.~\eqref{eq:ansatz}) to track the evolution of the normalized state vector. We record the evolution of the state vector norm using classical memory. As detailed in Appendix~\ref{app:norm_evo}, for an infinitesimal time step $dt$, the norm of $\ket{\tilde{\psi}\pqty{t}}$ shrinks according to
\begin{equation}
\label{eq:norm_evo}
\braket{\tilde\psi\pqty{t+dt}} \approx e^{- d \Gamma } \braket{\tilde\psi\pqty{t}} \ ,
\end{equation}
where $d \Gamma = 2\mel{\psi\pqty{t}}{H_\mathrm{a}}{\psi\pqty{t}} dt$, which is equivalent to $\mel{\phi}{H_\mathrm{a}}{\phi}$, can be measured using the same circuit that is used to measure the effective Hamiltonian. In Fig.~\ref{fig:bloch_sphere}, we provided a single qubit example using Bloch sphere to demonstrate of how our algorithm works.

\subsection{Restricted and unrestricted adaptive protocol}
%\textit{ Restricted and unrestricted adaptive protocol} -- 
Unlike existing variational methods~\cite{Benjamin_prl_2020} that use a fixed set of ansatz operators, our approach uses an adaptive procedure  that selects ansatz operators from a predefined pool during the time evolution.
The purpose of the adaptive protocol is to define a systematic way of preparing a compact ansatz. This is achieved by minimizing the McLachlan distance~\cite{MacDonald1934-mv} during each time step such that there are always enough operators in the ansatz to keep the distance below a threshold. However, a unique challenge for adaptive protocols is the lack of a known attainable lower bound for the McLachlan distance a priori, making it difficult to set a threshold value. This is an issue for both closed and open system adaptive algorithms. For the closed system case, if there are infinite operators in the pool, the lower bound should, in principle, be $0$. However, the same is not true for the open system case, which we will explain later.
To address this issue, we propose the unrestricted
adaptive variational quantum dynamics (UAVQDS) protocol, where instead of setting a fixed threshold for the McLachlan distance, we adopt a greedy approach which selects every operator that lowers the McLachlan distance.

 We formulate two versions of the adaptive protocol, the restricted version and unrestricted version. In the restricted version, the McLachlan's distance is set at a fixed threshold. While evolving, at every time step we measure the McLachlan's distance and a number bigger than the threshold triggers the adaptive procedure. We will call this the restricted adaptive variational quantum dynamics (RAVQDS).  On the other hand, an adaptive process can still be executed at every time step provided an operator in the operator pool lowers the McLachlan's distance. In other words, no matter what time step we are at, an operator will be added if it can lower the distance (by a relative threshold). Such a protocol ensures that the McLachlan's distance will reach the lowest possible value possible and not be restricted at a fixed threshold. We therefor call this the unrestricted adaptive variational quantum dynamics or UAVQDS.

In the generic case, the RAVQDS algorithm is not applicable to the open system case due to the lack of a reasonable threshold, even when there are infinite operators in the pool. We present a sufficient condition under which the RAVQDS could be applied, assuming the operators in the pool are expressive enough, and defer the proof to Appendix~\ref{app:lower_bound}.
A lower bound of the McLachlan distance square exists
\be
\mathcal{D} \ge 2\expval{\Ha^2} + 2\expval{\Ha}^2
\ee
if
\begin{subequations}
\label{eq:suff_cond}
\begin{align}
    \comm{\He}{\Ha} &= 0 \\
    \comm{A_\mu}{\Ha} &= 0 \ , \forall \mu \in \bqty{1, k} \ .
\end{align}
\end{subequations}
Eq.~\eqref{eq:suff_cond} seems to be a very strong condition. However, it is trivially satisfied when solving a quantum trajectory equation Eq.~\eqref{eq:unravelled_equation} with Lindblad operators as Pauli operators.

\subsection{Algorithms}
In this section we present the pseudocode of our algorithms. Algorithm~\ref{alg:vectorization} and Algorithm~\ref{alg:trajectory} show the vectorization method and the trajectory method respectively. The adaptive procedure in both these algorithms can be either restricted or unrestricted, depending on whether Eqs.~\eqref{eq:suff_cond} are satisfied. The pseudocode for the unrestricted procedure in presented in Algorithm~\ref{alg:unrestricted_adap}. In all the Algorithms, we use notations $\bth$ and $\mathbf{A}$ to denote the vector of the ansatz parameters (see Eq.~\eqref{eq:var_prin}) and the corresponding ansatz operators. Additionally, $\Gamma$ is the integrated norm shrinking factor given by $\Gamma \approx \sum_i d\Gamma_i$ where where $d\Gamma_i$ is the norm shrinking factor of the $i$th step (Eq.~\eqref{eq:norm_evo}).

\begin{algorithm}[ht]
\caption{Vectorization}\label{alg:vectorization}
\KwData{$ dt $\Comment*[r]{Step size} }
\KwData{$\ket{\psi(0)}$ \Comment*[r]{Initial state}}
\KwResult{$\bth,\ \mathbf{A},\ \Gamma$}
$t \gets 0$\;
$\Gamma \gets 0 $ \Comment*[r]{$\Gamma$ record the norm of $\ket{\tilde\psi\pqty{t}}$}
$\bth, \mathbf{A} \gets []^T$\Comment*[r]{Start with empty ansatz}
$\ket{\psi_\mathrm{R}}\gets\ket{\psi(0)}$\;
\While{$t < t_f$}{
    %Measure $M$ and $V$ in Eq.~\eqref{eq:theta_eom}\;
    update $\bth,\ \mathbf{A}, \ \dot{\bth}$ with the adaptive procedure\;
    $\bth \gets \bth + \dot{\bth} dt$ \; 
    $\Gamma \gets \Gamma + 2\expval{H_\mathrm{a}} dt$\;
    $t \gets t+dt$
}
\end{algorithm}

\begin{algorithm}[ht]
\caption{Trajectory}\label{alg:trajectory}
\KwData{$dt$ \Comment*[r]{Step size} }
\KwData{$\ket{\psi(0)}$ \Comment*[r]{Initial state}}
\KwResult{$\bth,\ \mathbf{A},\ \ket{\psi_\mathrm{R}}$} 
%\KwResult{$y = x^n$}
$t \gets 0$\;
$\Gamma \gets 0 $ \Comment*[r]{$\Gamma$ record the norm of $\ket{\tilde\psi\pqty{t}}$}
$\bth, \mathbf{A} \gets []^T$\Comment*[r]{Start with empty ansatz}
$\ket{\psi_\mathrm{R}}\gets\ket{\psi(0)}$\;
Generate a random number $q\in [0, 1)$ \;
\While{$t < t_f$}{
  \eIf{$e^{-\Gamma}\ge q$}{
    %Measure $M$ and $V$ in Eq.~\eqref{eq:theta_eom}\;
    update $\bth,\ \mathbf{A}, \ \dot{\bth}$ with the adaptive procedure\;
    $\bth \gets \bth + \dot{\bth} dt$ \; 
    $\Gamma \gets \Gamma + 2\expval{H_\mathrm{a}} dt$
  }{
      Randomly pick a jump operator $L_i$ according to the probability mass function given by Eq.~\eqref{eq:jump_p}\;
      $\ket{\psi_\mathrm{R}}\gets L_i\ket{\phi}/\norm{L_i\ket{\phi}}$\;
      $\bth, \mathbf{A} \gets []^T$\Comment*[r]{Reset the ansatz}
      $\Gamma \gets 0$\;
      Generate a random number $q\in [0, 1)$\;
  }
  $t \gets t+dt$
}
\end{algorithm}

\begin{algorithm}[ht]
\caption{Unrestricted adaptive procedure}\label{alg:unrestricted_adap}
\KwData{$\ket{\phi}$ \Comment*[r]{Parameterized state} }
\KwData{$r$ \Comment*[r]{Relative threshold} }
\KwResult{$\bth,\ \mathbf{A},\ \dot{\bth}$} 
%\KwResult{$y = x^n$}
Measure $\bM$ and $\bV$ in Eq.~\ref{eq:theta_eom}\;
$\dot\bth \gets \bM \backslash \bV$ \Comment*[r]{We use Tikhonov regularization for the pseudoinverse}
$\mathcal{D}'\gets\dot{\bth}^T \bM \dot{\bth}-2 \bV^T \dot{\bth}$\;
\Repeat{$\mathcal{D}' \ge \mathcal{D}-r$}{
    $\mathcal{D} \gets \mathcal{D}'$\;
    $\bM' \gets \bM, \ \bV' \gets \bV, \ \bth' \gets \bth$\;
    \ForEach{$A_k$ in the operator pool}{
        update $\bM''$ and $\bV''$ \Comment*[r]{Add $A_k$ and $\theta_k\equiv 0$ to the ans\"atze}
        $\dot\bth'' \gets \bM'' \backslash \bV''$\;
        $\mathcal{D}''\gets\dot{\bth''}^T \bM'' \dot{\bth''}-2 \bV'^T \dot{\bth''}$\;
        \If{$\mathcal{D}''<\mathcal{D}'$}{
            $\mathcal{D}' \gets \mathcal{D}''$\;
            $\bM'\gets\bM'',\ \bV'\gets\bV'', \ \bth'\gets \bth'', \ \dot{\bth}'\gets \dot{\bth}''$
        }
    }
}
$\bth\gets \bth',\ \mathbf{A}\gets \mathbf{A}',\ \dot{\bth}\gets \dot{\bth}'$\;
\end{algorithm}

Before presenting the simulation results, we would like to make a few comments on the algorithms used. In the vectorization method, we need to update $\bth$ and $\mathbf{A}$ at each time step and keep track of the norm $e^{-\Gamma}$. The final vectorized state is $e^{-\Gamma/2}\ket{\phi\pqty{t_f}}$. However, to measure any observable $O$ in the unvectorized setting, we will need to evaluate 
\begin{equation}
    \expval{O} = e^{-\frac{\Gamma}{2}}\norm{O}_1\braket{O^\dagger}{\phi(t_f)}
\end{equation}
%$e^{-\Gamma/2}\braket{O^\dagger}{\phi(t_f)}$ (Eq.~\eqref{eq:iso_trace}) 
where $\ket{O^\dagger}$ is the normalized and vectorized operator $O^\dagger$ and $\norm{\cdot}_1$ denotes the trace norm.  This quantity can be measured either directly or indirectly with the help of a synthesized unitary $V$ which prepares $\ket{O^\dagger}$, i.e. $\ket{O^\dagger}=V\ket{0}^{\otimes 2N}$~\cite{Kamakari2022-yr} (See Appendix~\ref{app:mO} for details).

In the trajectory method, the total evolution is divided into intervals of a parameterized circuit followed by a jump operator. Each interval can be regarded as the state preparation circuit for the next one. As a consequence, the circuit depth of this approach increases with the number of jumps. Because we only need to measure the normalized state vector in this case, any measurement circuit described in Appendix~\ref{app:MV_measurement} can be used. 
%\hl{In this scenario, we define the expectation value of an observable $O$ for $N_{\tau}$ trajectories as}
%\be 
%\ev{O(t)} = \frac{1}{N_{\tau}}\sum_{\lambda=1}^{N_{\tau}} \expval{O}{\phi^{(\lambda)}(t)} 
%\label{eq:trajectory_expectation}
%\ee 

Lastly, a threshold $r$ which specifies the minimum amount reduction in the McLachlan distance needed to keep the adaptive procedure running is still necessary in the unrestricted algorithm. It can be either a additive threshold or a multiplicative threshold. In this paper, we will use the additive threshold as described in Algorithm~\ref{alg:unrestricted_adap}, and refer it as the adaptive threshold. In addition, the algorithms (restricted or unrestricted ) require us to solve the linear equations given by Eq.~\eqref{eq:theta_eom}, which can become ill-conditioned as the size of ans\"atze increases. To address this challenge, Tikhonov ($L_2$) regularization
\begin{equation}
    \label{eq:tik_reg}
    \dot{\bth} = \pqty{\bM^\dagger \bM + \lambda \mathbf{I}}^{-1}\bM^\dagger \bV \ ,
\end{equation}
is applied when inverting $\bM$. Here $\lambda$ is a small parameter to shift the diagonals of the $\bM^\dagger \bM$ matrix. 
\section{Result}
\subsection{Noiseless simulation}
\label{sec:noiseless_dim}
The QA simulation is carried out by evolving a system from $t=0$ to $t=t_f$ under a time-dependent  Hamiltonian of the form
\be
    H\pqty{t} = A\pqty{t}H_\mathrm{D} + B\pqty{t} H_{\mathrm{P}} \ ,
    \label{eq:hamit_t}
\ee
where $A\pqty{t}$ and $B\pqty{t}$ (known as the annealing schedule) are scalar functions which satisfy $A\pqty{0} \gg B\pqty{0} \approx 0$ and $B\pqty{t_f} \gg A\pqty{t_f} \approx 0$. Here, we will only consider the linear schedule
\be
A\pqty{t} = 1-t/t_f \ , \quad B\pqty{t} = t / t_f \ .
\ee
%For the $N$ qubit-case, we study the alternating sector chain (ASC) problem. It is a well-understood $1$-D chain problem and can serve as a benchmarking problem for our algorithm. The total Hamiltonian of the system is
$H_\mathrm{D}$ and $H_\mathrm{P}$ are constant terms given by
\begin{equation}
    \label{eq:hd_hp}
    H_\mathrm{D} = -\sum_{i=1}^N X_i \ , \quad H_\mathrm{P} = -\sum_{i=1}^{N-1}j_iZ_iZ_{i+1} \ ,
\end{equation}
where $N$ is the total number of qubits and $Z_i$ ($X_i$) is the Pauli $Z$ ($X$) matrix on the $i$th qubit. The couplings strength $j_i$s are given by
\begin{equation}
    j_{i}= \begin{cases}w_{1} & \text { if }\lceil i / n\rceil \text { is odd } \\ w_{2} & \text { otherwise }\end{cases} \ ,
\end{equation}
where $n$ is the sector size. In this study, we fix the model parameters as $w_1 = 1$, $w_2 = 0.5$ and $n=1$.

We focus on two types of Lindblad models, with the first one consisting of only $Z_i$ Lindblad operators applied to each qubit, given by:
\be
    L_i = Z_i \ , \quad i \in \bqty{1, N}
\ee
with rate $\gamma_i$. It describes a continuous dephasing channel on each qubit. It is worth noting that the conditions outlined in Eqs.~\eqref{eq:suff_cond} are satisfied for the corresponding unravelled equation in this case. Therefore, the RAVQDS method can be applied for the trajectory method. We will refer to this model as the dephasing model going forward, and we will fix the values of dephasing rates as $\gamma_i = 0.01$ in our simulations.
%For the pure dephasing model, the  the anti-Hermitian in the effective Hamiltonian (Eq.~\eqref{eq:split_H}) 
The second model consists of two Lindblad operators on each qubit
\be
    L_i^+ = \left( X_i + i Y_i \right) /2 \ ,\quad L_i^- = \left( X_i - i Y_i \right) /2
\ee
with different rate $\gamma_i^+$ and $\gamma_i^-$. This model is often referred to as an empirical model for an incoherent energy transfer process and can also be rigorously derived from  first principles. We will refer to this model as the amplitude damping model, and fix the values of the rates as $\gamma_i^+ = 0.04$ and $\gamma_i^- = 0.004$ henceforth. It is worth noting that we choose the $\gamma$ values based on two considerations: First, they should be small enough to ensure the weak coupling limit still holds; Second, they need to be large enough to make the open system effects non-negligible at the time scale of our simulation. An additional challenge arises when implementing the trajectory simulation of the amplitude damping model, because the jump operators $L^+_i$ and $L^-_i$ are non-unitary. Either mid-circuit measurement or block encoding (unitary dilation)~\cite{Low2017-ye,Del_Re2020-ul,camps2023explicit} based methods can be adopted to implement those non-unitary jumps (See Appendix~\ref{app:jumps} for details).

We conducted numerical experiments to compare different operator pools.
The choice of our operator pool is a combination of the Hamiltonian pool~\cite{Yao2021-is} and the qubit-adapt pool~\cite{vqe_qubit_adaptive}. Initially, we utilize the smallest pool, namely the Hamiltonian pool, and incrementally enlarge it if the algorithm does not achieve the desired level of accuracy. While there can be multiple choices of operator pools, variational ans\"atze generated with qubit-adapt pools are much shallower~\cite{vqe_qubit_adaptive, AVQITE}. Starting from a pool consisting of all single qubit Pauli operators 
$\mathcal{P}_{s} =\{X_i\}_{i=1}^{N'}\cup\{Y_i\}_{i=1}^{N'} \cup\{Z_i\}_{i=1}^{N'}$, we constructed three distinct operator pools by adding three different types of two-qubit Pauli operators to $\mathcal{P}_{s}$, defined as follows:
\begin{subequations}
\label{eq:pool_eq}
    \begin{align}
        \mathcal{P}_{1} &= \mathcal{P}_{s}\cup \{Z_{i}Z_{i+1}\}_{i=1}^{N'-1}\\
         \mathcal{P}_{2} &= \mathcal{P}_{s}\cup \{P_{i}P_{i+1}\}_{i=1}^{N'-1} \\
         \mathcal{P}_{3} &= \mathcal{P}_{s}\cup \{P_{i}P_{j}\}_{i,j=1}^{N'} \ ,
    \end{align}
\end{subequations}
where $P_{i} \in \{ X_i, Y_i, Z_i\}_{i=1}^{N'}$ and $N'$ denotes the number of physical qubits.  We use the individual terms from Eqs.~\eqref{eq:pool_eq} as the operators in our pool.  $\mathcal{P}_{1}$ draws its inspiration from the Hamiltonian pool (the terms in the Hamiltonian appear in the pool only). 
We found the $\mathcal{P}_{1}$ fails to recover the desired dynamics, which could be because a Hamiltonian pool is effective only for simulating unitary dynamics.  Since the problem of open quantum system deals with non-unitary components as well, a Hamiltonian pool is insufficient in this case. While $\mathcal{P}_{2}$ generates the desired results for the trajectory method, it fails in the vectorization method case, which could be attributed to the non-local term $\bar{L}_k\otimes L_k$ in the vectorized effective Hamiltonian (Eq.~\eqref{eq:Heff_vec}). Only $\mathcal{P}_{3}$ proves effective in producing the desired results for both methods. We summarize the pool dependence of our method in Table~\ref{table:1}.
\begin{table}[!ht]
\centering
\begin{tabular}{||c | c | c | c||} 
 \hline
\diagbox{Method}{Pool} & $\mathcal{P}_1$ & $\mathcal{P}_2$ & $\mathcal{P}_3$ \\  
 \hline\hline
 Trajectory & $\times$ & Yes & Yes \\
 Vectorization & $\times$ & $\times$ & Yes \\
 \hline
\end{tabular}
\caption{\textbf{Dependence of our method on operator pools.}}
\label{table:1}
\end{table}

%Finally, we conducted numerical experiments to compare different operator pools. All the pools included every single qubit Pauli operator on each qubit (we will omit this part later when refer to different pools), but differed in the choices of two-qubit Pauli operators. Specifically, we considered pools with $ZZ$ operators between neighboring qubits, every two-qubit Pauli operator between neighboring two qubits, and every two-qubit Pauli operator between each pair of qubits. 
%We found that the pool with only nearest neighbor $ZZ$ operators did not work for either the vectorization or the trajectory method. On the other hand, the pool with only nearest neighbor Pauli operators worked only for the trajectory method, while the pool with every two-qubit Pauli operator worked for both methods.
%\HC{For vectorization method, we also try the $P_i\otimes P_{i+N}$.}
%\begin{figure}[t]
%    %\centering
%    \includegraphics[width=\linewidth]{figure/fig1.png}
%    \caption{ \textbf{Noiseless simulation using UAVQDS:}  Energy and population evolution using for model 1 ((a) and (b)) and model 2 ((c) and (d)) respectively, for  $N=8$. }
%    \label{fig:n8_results}
    %source files: MacAir: /Users/niladri/Documents/Hubbard/param_ground_state/ground_state/gs_plot6.py
%\end{figure}
\begin{figure*}[htb]
    \centering
    \subfloat[]{
    \includegraphics[width=0.47\linewidth]{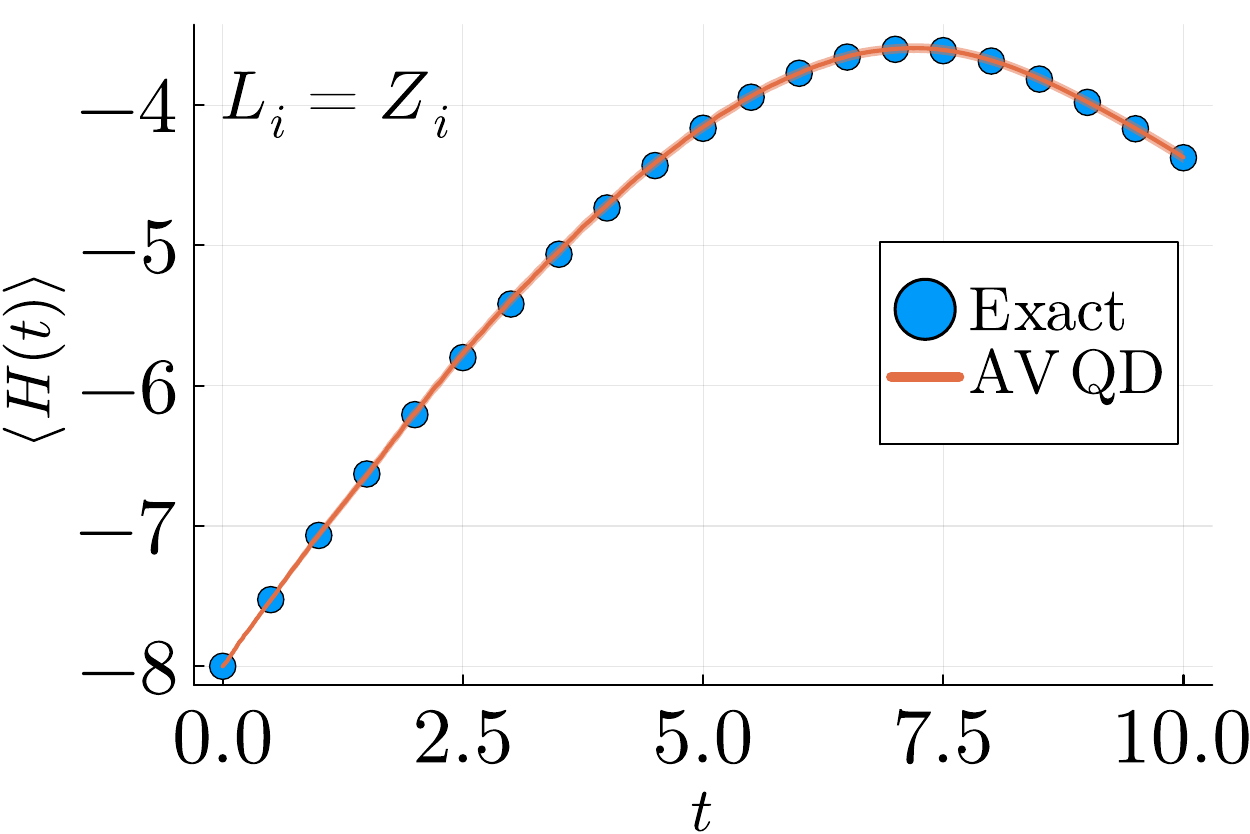}
    }
    \hfill
    \subfloat[]{%
    \includegraphics[width=0.47\linewidth]{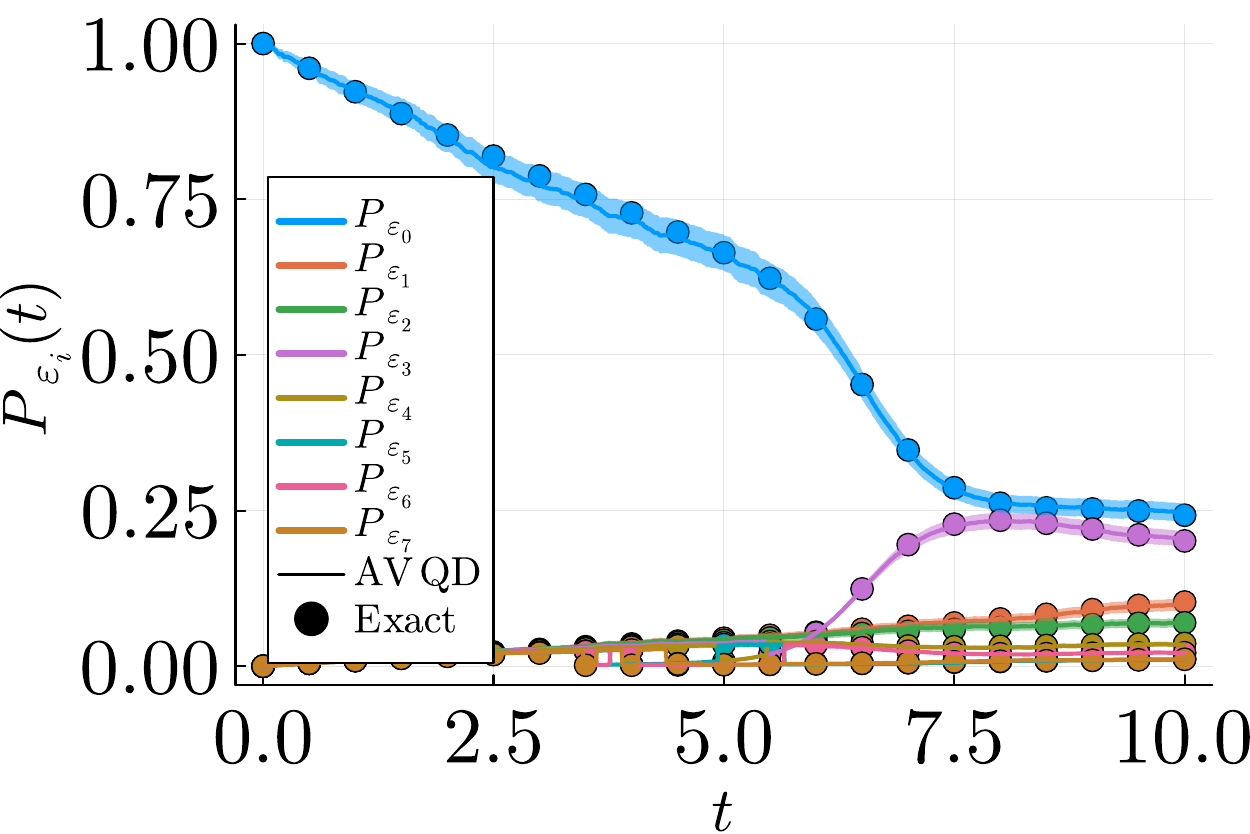}
    }%
    \\
    \subfloat[]{
    \includegraphics[width=0.47\linewidth]{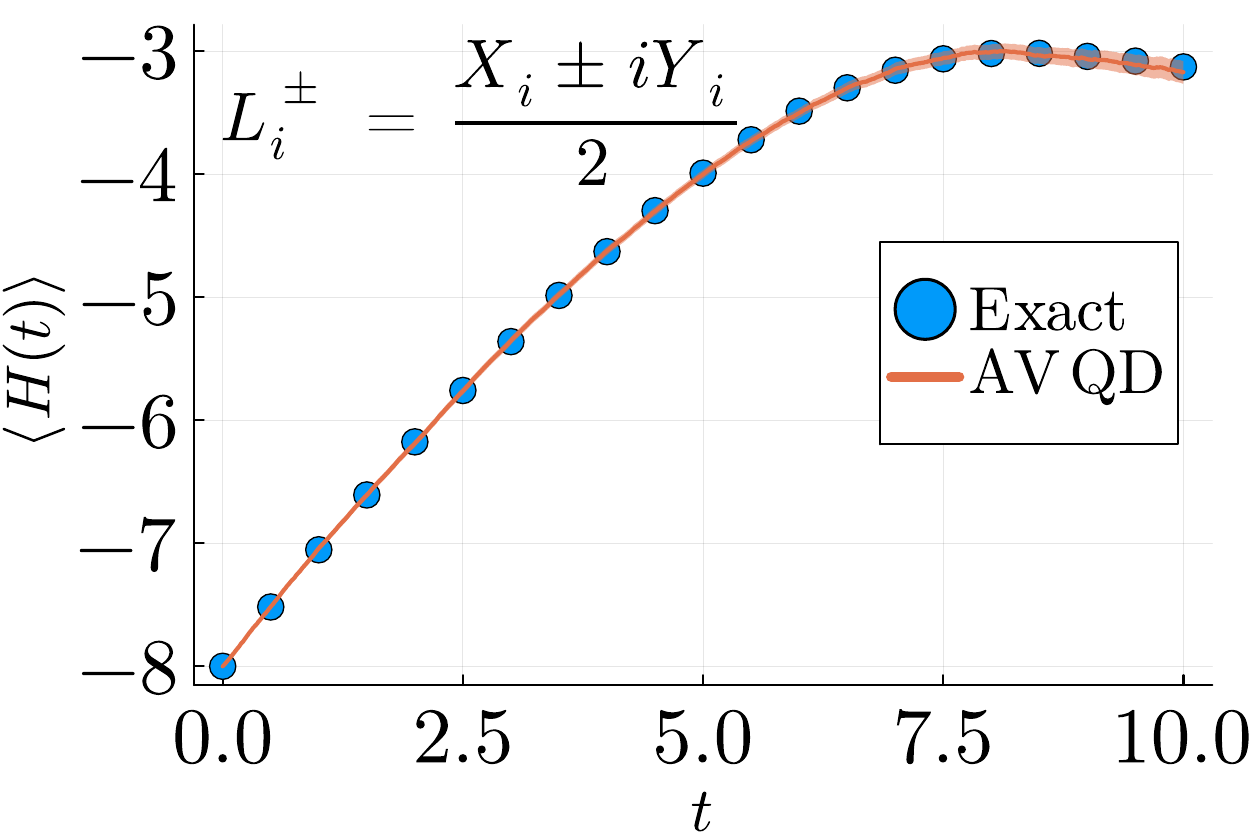}
    }
    \hfill
    \subfloat[]{
    \includegraphics[width=0.47\linewidth]{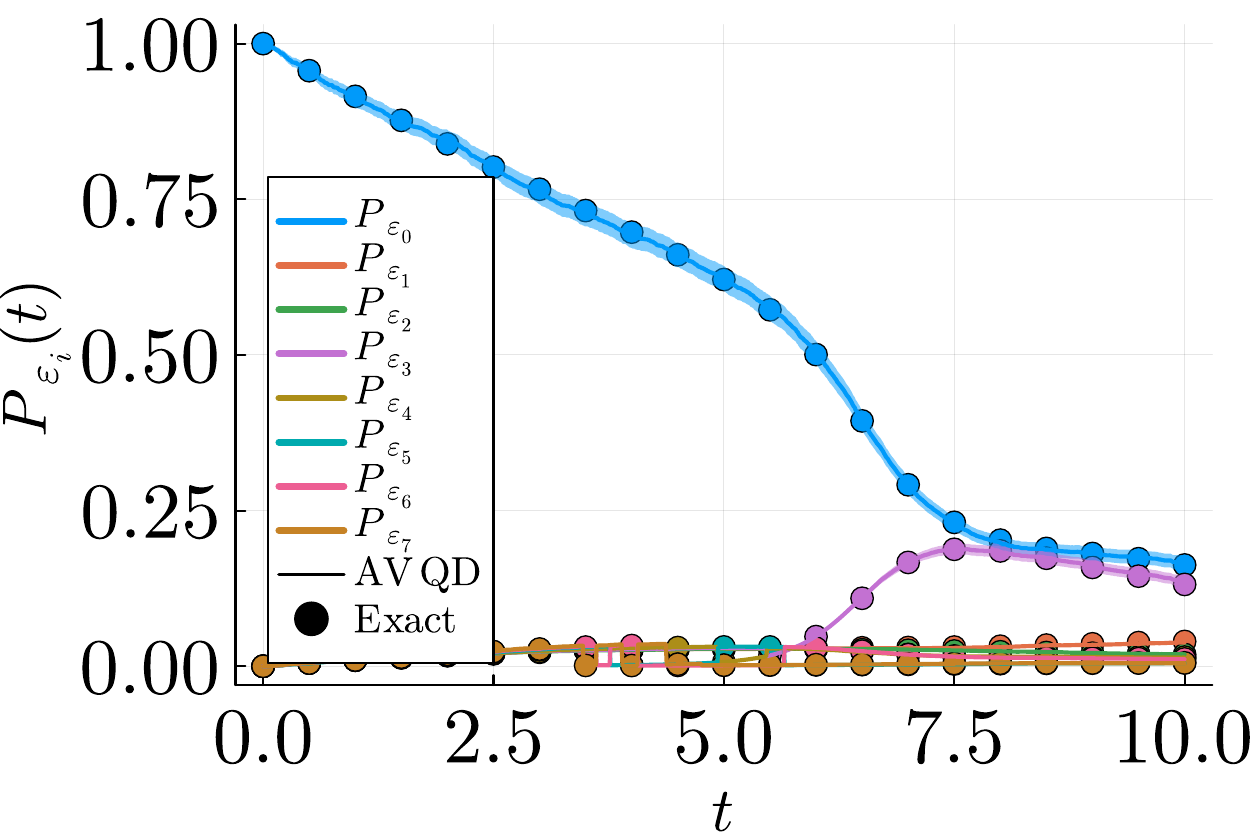}
    }%
    \caption{ \textbf{Noiseless simulation using the trajectory method and UAVQDS for  $\mathbf{N=8}$ ($\mathbf{8}$ physical qubits)}. Energy and eigenstate population evolution using the dephasing model ((a) and (b)) and the amplitude damping model ((c) and (d)) are shown respectively. The UAVQDS results are obtained from $1000$ trajectories and the ribbons represent the $2\sigma$ ($\sigma$ is the standard error of the mean) error bar for the trajectory average. The adaptive threshold was set to $r=10^{-4}$ and the time step size was set to $dt=0.01$. We used the operator pool $\mathcal{P}_2$.}
    \label{fig:n8_results_trajectory}
\end{figure*}

\begin{figure*}[htb]
    \centering
    \subfloat[]{
    \label{fig:4qubit_Z_vec}
    \includegraphics[width=0.47\linewidth]{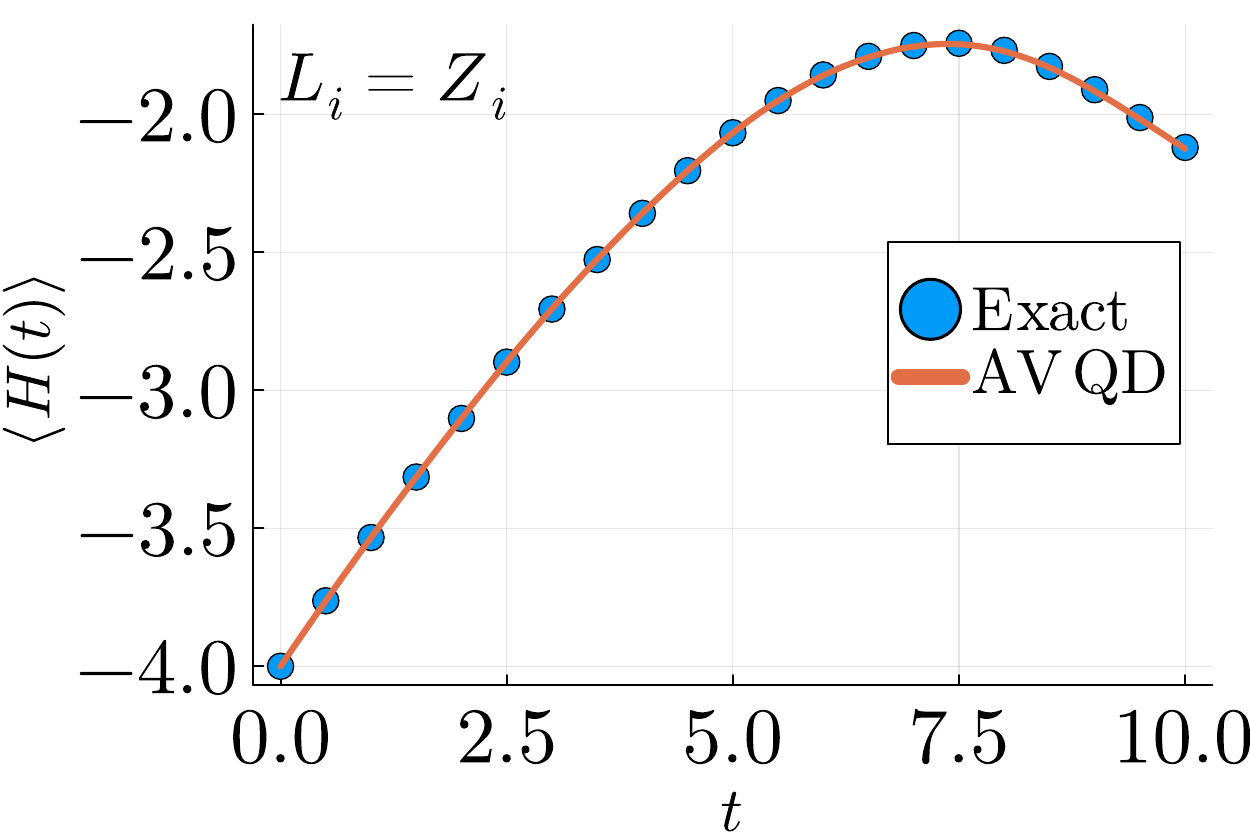}
    }
    \hfill
    \subfloat[]{%
    \includegraphics[width=0.47\linewidth]{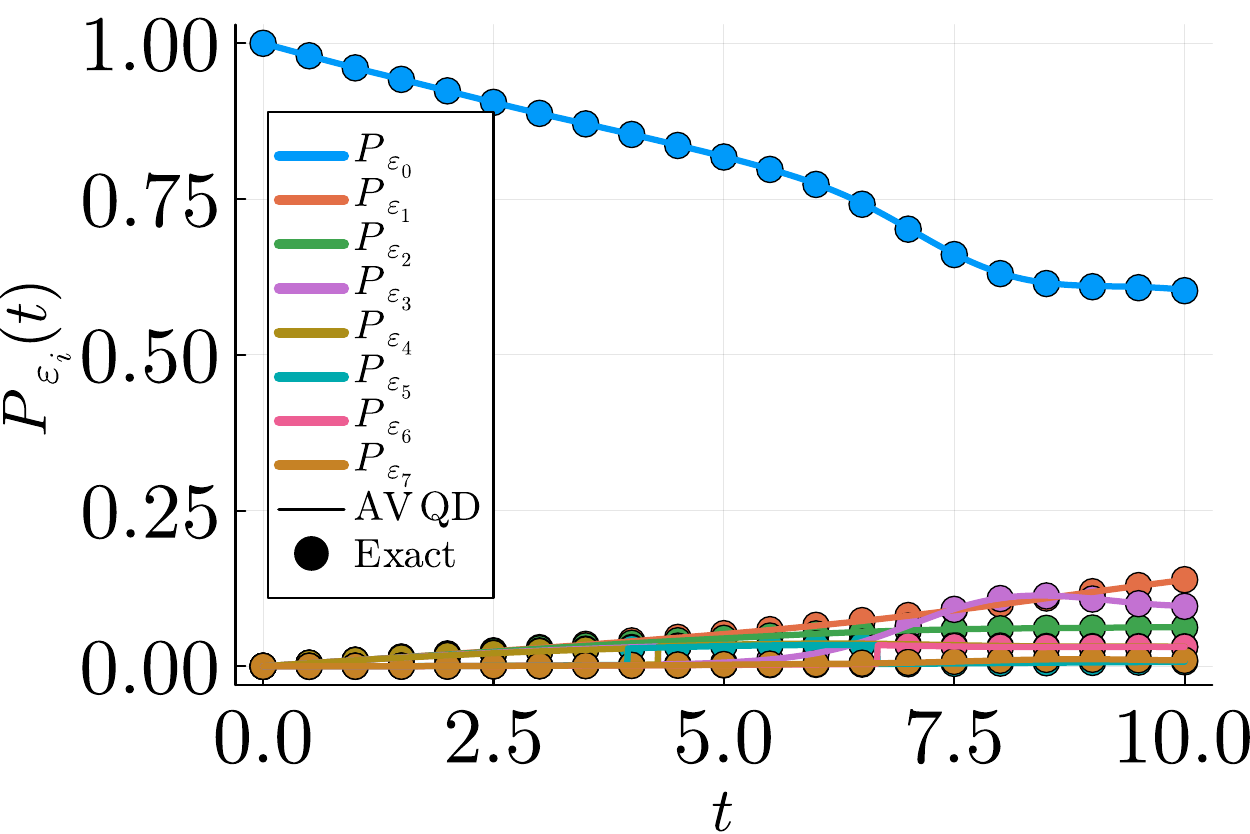}
    }%
    \\
    \subfloat[]{
    \label{fig:err_damping}
    \includegraphics[width=0.47\linewidth]{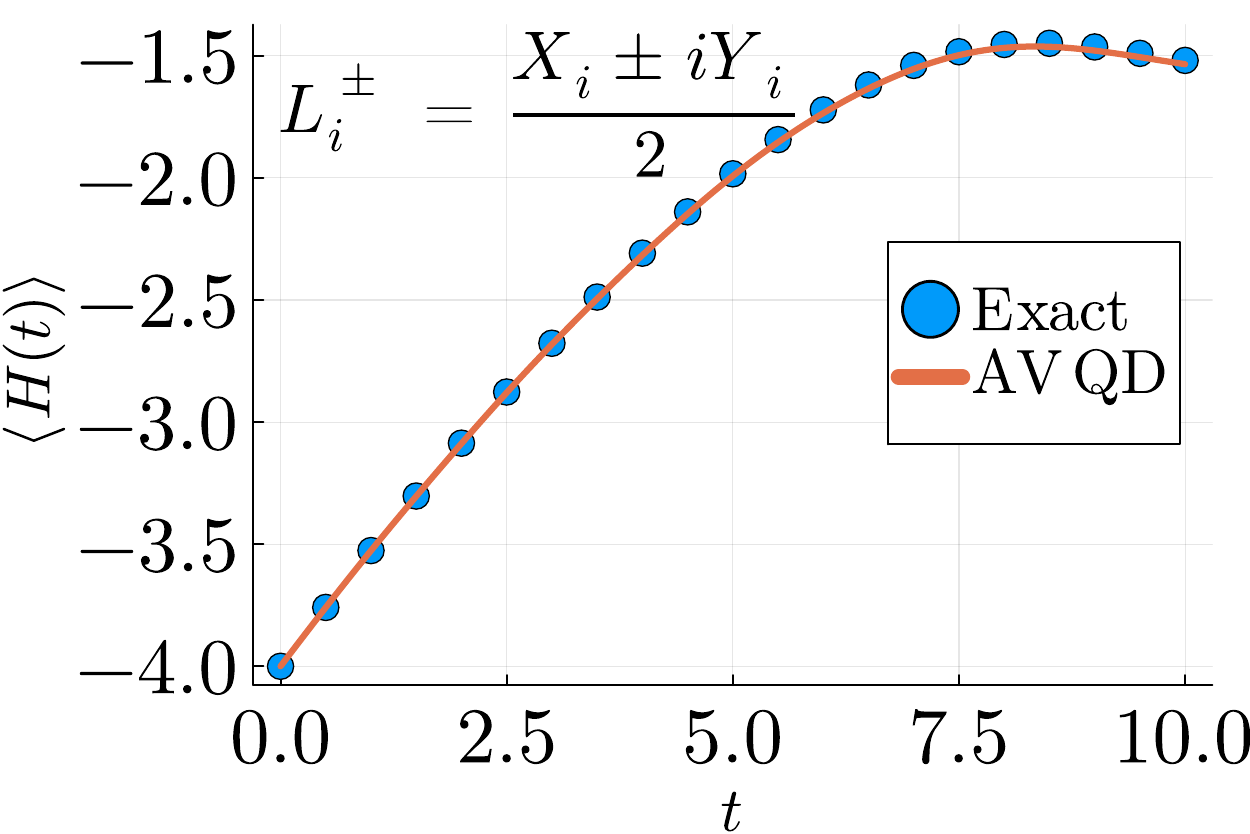}
    }
    \hfill
    \subfloat[]{
    \label{fig:pop_damping}
    \includegraphics[width=0.47\linewidth]{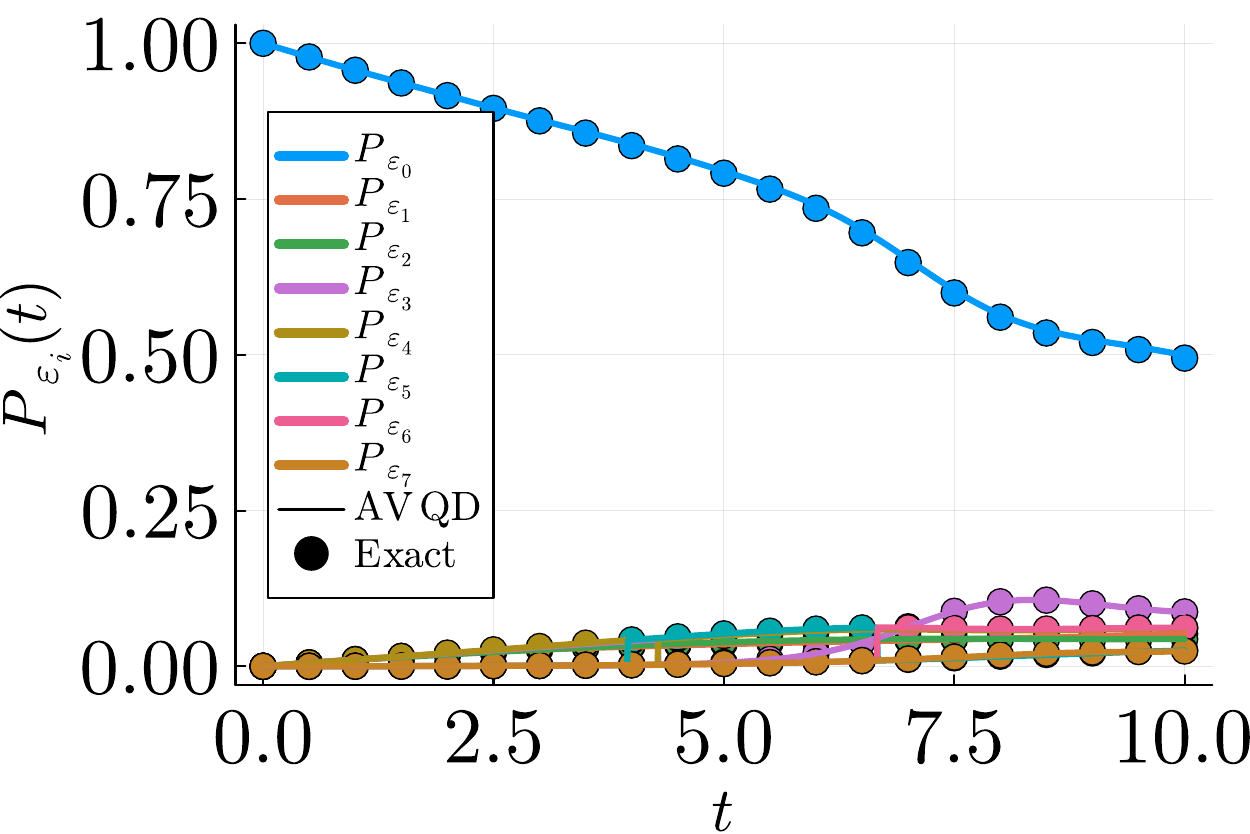}
    }%
    \caption{\textbf{Noiseless simulation using the vectorization method and UAVQDS for  $\mathbf{N=4}$ ($\mathbf{8}$ physical qubit)}. Energy and eigenstate population evolution using the dephasing model ((a) and (b)) and the amplitude damping model ((c) and (d)) are shown respectively. For the dephasing model, the adaptive threshold was set to $r=10^{-6}$, and for the amplitude damping model, it was set to $r=10^{-7}$. The time step size was fixed at $dt=0.01$. The operator pool $\mathcal{P}_3$ was used for these simulations.}
    \label{fig:n8_results_vectorization}
\end{figure*}

%{\it{Operator pool completeness}}- 

When designing an operator pool, an important consideration is identifying the minimal complete pool (MCP), which minimizes the extra measurements needed for the adaptive process. MCP has been extensively studied in the case when the state vector is real~\cite{vqe_qubit_adaptive, Shkolnikov2023-kg}, where an explicit MCP is suggested. However, for the problem we study in this paper (Eq.~\eqref{eq:hd_hp}), the state vector is not limited to real vectors. As a result, the MCP proposed in the aforementioned references is are not directly applicable to our case. Although we can reformulate the problem such that the existing MCP results would apply, i.e., the state vector is always real, by expressing Eq.~\eqref{eq:Lindblad_eq} in the Pauli basis, we have opted not to pursue this route in our current study. This is because such a representation would lead to an effective Hamiltonian comprising 4-body Pauli terms, potentially increasing circuit complexity (see Appendix~\ref{appendix:vectorization_pauli_basis} for a detailed discussion). Instead we provide an upper bound on the number of operators necessary for an MCP applicable to any state vector by selecting the operator pool such that it can generate a universal gate set. One such pool comprises every single Pauli operator and the $ZX$ operator on adjacent qubits, leading to an MCP size upper bound of $4n-1$. On the other hand, as indicated by~\cite{Shkolnikov2023-kg} and by our numerical findings presented in Table~\ref{table:1} ($\mathcal{P}_2$ is already a complete pool), having a complete pool does not guarantee optimal outcomes. Investigating strategies that enhance the algorithm's ability to converge towards accurate results represents a promising direction for further research.

%An operator pool $\mathcal{P_C}$ with a set of operators $\{A_{\mu}\}$ is called compete if for any quantum state $\ket{\psi}$, the rank $(r)$ of the matrix $\mathcal{M}_{\mu \nu} = \ev{A_{\mu}A_{\mu}}{\psi}$ is greater than or equal to $2^{n} - 1$, where $n$ is the number of qubits \cite{vqe_qubit_adaptive, Shkolnikov2023-kg}. There are minimal pools that are  provably complete, where the operators are sufficient to rotate any real state to
%any other real state in the Hilbert space. We tested our algorithm with the minimal complete pool as defined in \cite{vqe_qubit_adaptive}, but failed to reach the desired convergence. However, by using the pool $\mathcal{P}_2$ and $\mathcal{P}_3$, which are supersets of the complete operator pool, a satisfactory algorithmic accuracy could be reached (see Table~\ref{table:1}). This implies that the minimal complete operator pool for open systems differs from that of closed systems. The search for such a pool can be numerically extensive and also require an analytical approach. We aim to investigate this in our upcoming work. 

In Figs.~\ref{fig:n8_results_trajectory} and~\ref{fig:n8_results_vectorization}, we present the results from noiseless simulations of running UAVQDS with both the trajectory and vectorization methods on the dephasing and amplitude damping models. We compare the evolution of the energy $\ev{H(t)} = \Tr\{\rho H(t)\}$ and instantaneous eigenstate populations $P_{\varepsilon_{i}}(t) = \expval{H(t)}{\varepsilon_i\pqty{t}} $ obtained from UAVQDS with the exact solution obtained from the Hamiltonian open quantum system toolkit (HOQST) package~\cite{Chen2022-ps}. Here, $\ket{\varepsilon_i(t)}$ represents the $i$th instantaneous eigenstate of the Hamiltonian, i.e., $H(t) \ket{\varepsilon_i\pqty{t}} = E_i\ket{\varepsilon_i\pqty{t}}$.
%\hl{The expectation value} $\ev{H(t)} $
%\hl{in the trajectory method is calculated according to eq.}~\eqref{eq:trajectory_expectation}. The population \hl{The population is similarly calculated as} $E_i = $. 
The eigenstate populations are chosen as benchmarks because they are the quantities of interest in the original QA setting. The non-vanishing excited-state populations indicate that the dynamics happens in the non-adiabatic regime, where quantum effects are expected to play a non-trivial role~\cite{Dickson2013-kf, Boixo2016-oy, Crosson2021-yg, Garcia-Pintos2023-lz, Munoz-Bauza2019-ad}. Since the primary focus of our work in not QA, we direct readers keen on further details to the references mentioned above. 

While it is still unclear whether the eigenstate populations can be efficiently measured on a real quantum computer, they can be easily obtained in the simulation. We will focus only on the energy evolution when implementing the algorithm on the hardware. When running the simulations, we choose a stepsize of $dt=0.01$ with a total evolution time $t_f = 10$. Additionally, we set the Tikhonov regularization parameter $\lambda$ in Eq.~\eqref{eq:tik_reg} to a fixed value of $10^{-8}$. Before proceeding, it's important to note that the instantaneous eigenstate populations depicted in Figs.~\ref{fig:n8_results_trajectory} and~\ref{fig:n8_results_vectorization} do not sum to one as they represent only the lowest 
$8$ energy states. We have confirmed that the trace of the density matrix throughout the evolution remains unity. %$1$.

%\begin{itemize}
%    \color{red}
%    \item The results agree with the exact solution.
%    \item The scaling is not exponential with system size.
%    \item The scaling is not exponential with error.
%\end{itemize}

%Remarkably, our results show that UAVQDS accurately captures the dynamics of the high energy levels suggesting its potential utility in exploring quantum dynamics that are challenging to simulate classically.
The figures demonstrate that UAVQDS produces results in good agreement with the exact solver, as the dots representing the exact solutions overlap with the UAVQDS lines across all plots. It is noteworthy that compared to the trajectory method, the vectorization method requires double the number of qubits and a bigger operator pool size with non-nearest neighbor coupling. Hence, the ansatz there is more complex and more expensive to generate.  For instance, $r = 10^{-4}$ was used to run the ($N=8$) trajectory problem (with $8$ physical qubits), whereas for the ($N=4$) vectorized problem (also with $8$ physical qubits), $r$ had to be reduced to $10^{-6}$ ($10^{-7}$ for the amplitude damping model)  to ensure enough operators were included in the simulation (See Appendix~\ref{app:vec_figs} for more data).
%However, it is important to note that achieving similar accuracy with the trajectory method requires a much lower adaptive threshold and a larger operator pool in the vectorization method. 
%For Fig.~\ref{fig:n8_results_vectorization}, we used an operator pool that includes every 2-qubit-Pauli operator, as opposed to the neighboring 2-qubit-Paulis used for Fig.~\ref{fig:n8_results_trajectory}.
%We observed that the error for the vectorized case in Fig.~\ref{fig:err_damping} is larger than the other cases, despite having a smaller $N$. 
This implies that the effective Schr\"odinger equation generated by the vectorization method poses a greater challenge for the UAVQDS than the one produced by the trajectory method. This introduces an additional trade-off between the vectorization and trajectory methods, in addition to computational space (number of qubits) and execution time (number of trajectories).

%\begin{figure}[t]
    %\centering
%    \includegraphics[width=\linewidth]{figure/fig2.png}
%    \caption{ \textbf{Operators and CNOT usage in UAVQDS simulation:} (a) Number of operators for each trajectory for $N=8$. Each bar represents the number of operators in the ansatz. Different colors in each bar depicts a jump.  (b) Scaling of the average number of operators with system size}
%    \label{fig:scaling}
    %source files: MacAir: /Users/niladri/Documents/Hubbard/param_ground_state/ground_state/gs_plot6.py
%\end{figure}

\begin{figure*}[htb]
    \centering
    \subfloat[]{
    \label{fig:err_r}
    \includegraphics[width=0.47\linewidth]{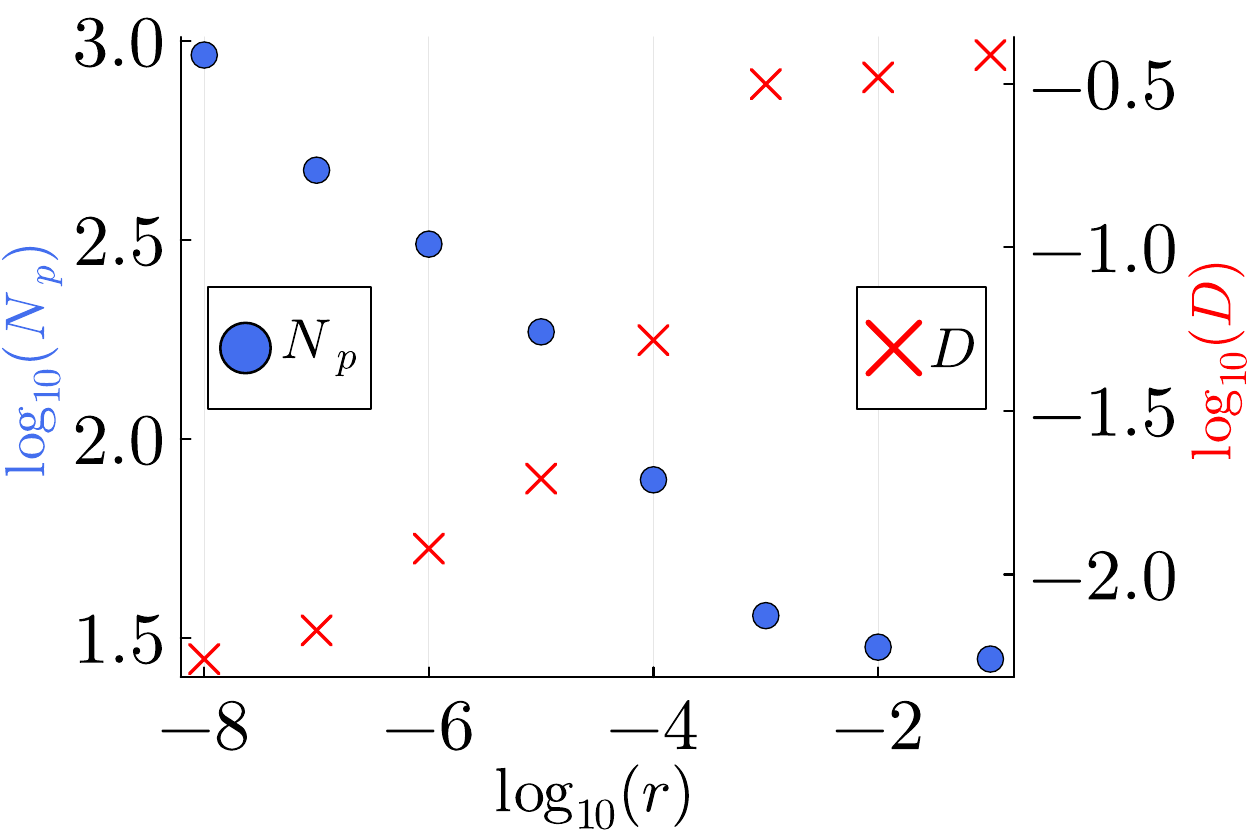}
    }
    \hfill
    \subfloat[]{%
    \label{fig:err_n}
    \includegraphics[width=0.47\linewidth]{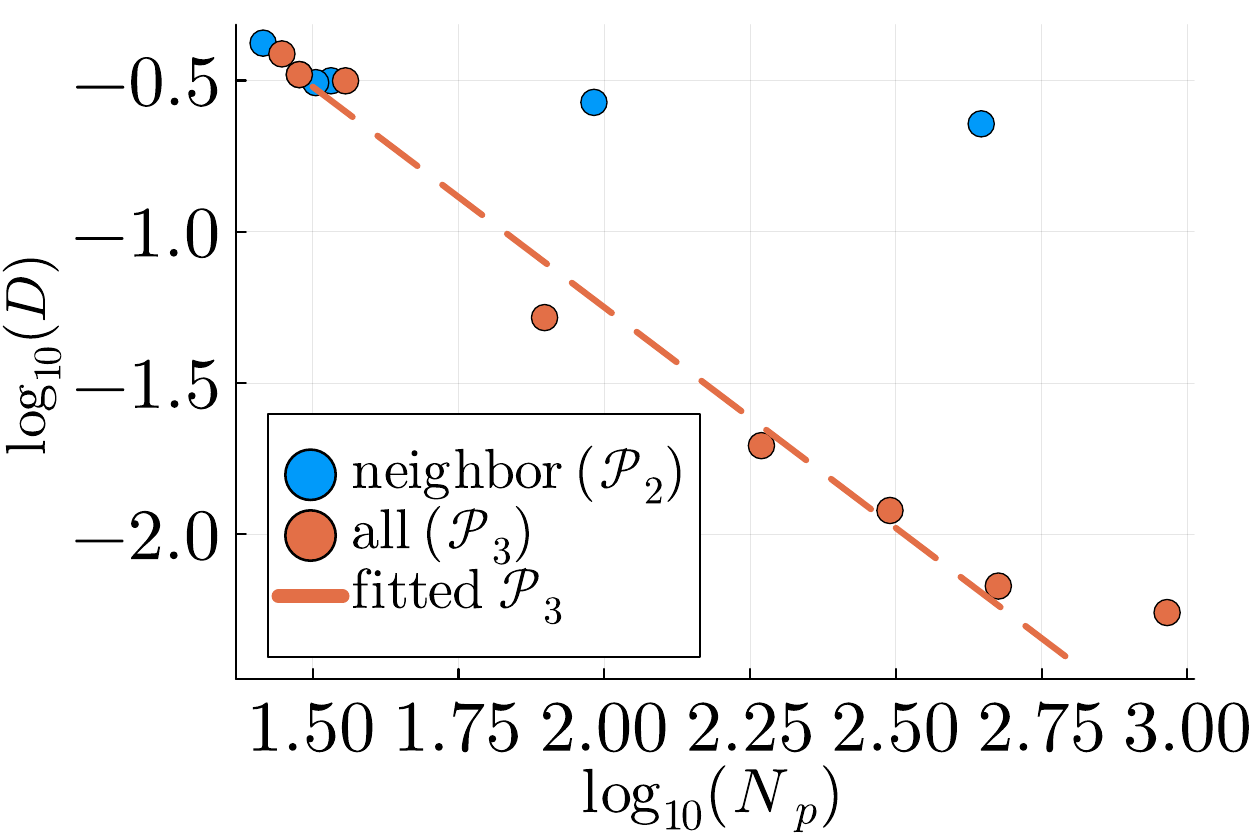}
    }%
    \caption{ \textbf{Infidelity of UAVQDS versus adaptive threshold and total number of parameters in the ansatz.} Panel (a) shows the total number of parameter in the ans\"atze at the end of evolution (blue dots) and the infidelity (red crosses) versus the adaptive threshold $r$. The operator pool employed includes all 2-qubit Paulis ($\mathcal{P}_3$). Panel (b) shows the infidelity versus the total number of parameters in the ans\"atze for pools with either neighboring (blue dots) or all (orange dots) 2-qubit Paulis ($\mathcal{P}_2$ and $\mathcal{P}_3$, respectively). The dashed line in (b) represents a fitted line using all data from the orange dots except the last one. The results were obtained using the setup from Fig.~\ref{fig:4qubit_Z_vec}.}
    \label{fig:n4_err_scaling}
\end{figure*}

%\begin{figure}[htb]
%    \centering
%    \subfloat[]{
%    \includegraphics[width=0.47\linewidth]{figure/ASC_8_qubit_lind_Z_rcut_1.00e-04_unrestricted_parameters_num.pdf}
%    }
%    \hfill
%    \subfloat[]{%
%    \includegraphics[width=0.47\linewidth]{figure/ASC_lind_Z_rcut_1.00e-04_unrestricted_scaling.pdf}
%    }%
%    \caption{Operators and CNOT usage in the trajectory-method-based UAVQDS simulation. (a) Number of ans\"atze versus the evolution time for three different trajectories. The vertical drop indicates a jump event. All the trajectories have the same parameters until the first jump. (b) Box plot of the cumulative number of ans\"atze and estimated CNOT-gate count versus the system size. The CNOT-gate count is estimated using two times the number of 2-qubit ansatz operators. The simulation parameters are the same as those used in Fig.~\ref{fig:n8_results_trajectory}.}
%    \label{fig:n8_Z_params}
%\end{figure}
We further investigate the error of the vectorization UAVQDS in Fig.~\ref{fig:n4_err_scaling}. To quantify the error, we define an infidelity measure $D$ as the trace-norm distance between the solution of UAVQDS and the exact solution at the end of the evolution, given by
\begin{equation}
    D = \Tr \bqty{\sqrt{\pqty{\rho(t_f) - \rho_t}^\dagger\pqty{\rho(t_f)-\rho_t}}} \ ,
\end{equation}
where $\rho(t_f)$ and $\rho_t$ represent the solutions obtained from UAVQDS and the exact solver, respectively. As shown in Fig.~\ref{fig:err_r}, by lowering the adaptive threshold, we increase the accuracy of the algorithm at the cost of more ansatz operators. However, this trend will not continue indefinitely because there are only a finite number of operators in the predefined pool. The relationship between the infidelity and total number of parameters in the ans\"atze is illustrated in Fig.~\ref{fig:err_n}. The infidelity decreases approximately polynominally (linearly in $\log$-$\log$ plot) as more operators are included in the ans\"atze until it hit a flat region. To further reduce the infidelity, a larger pool is required.

\begin{figure*}[htb]
    \centering
    \subfloat[]{
    \label{fig:nscaling_t1}
    \includegraphics[width=0.47\linewidth]{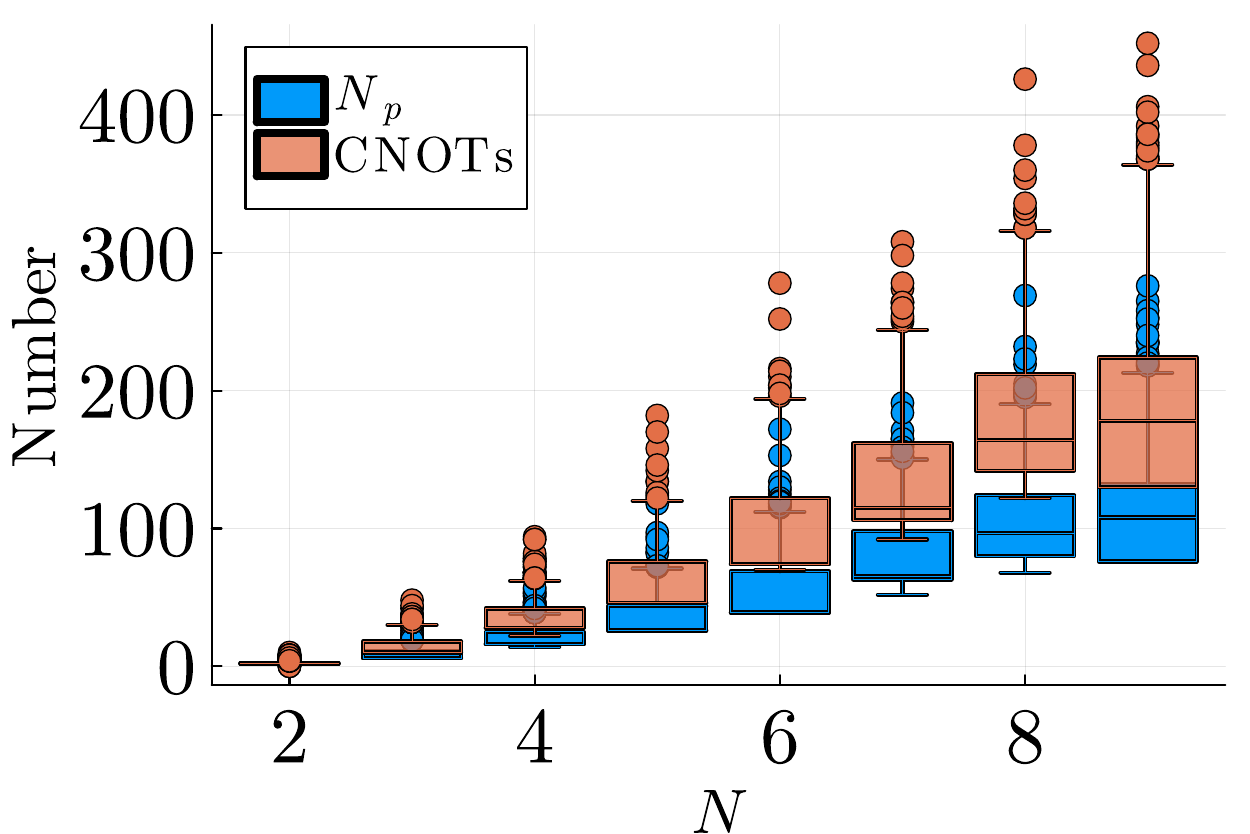}
    }
    \hfill
    \subfloat[]{%
    \label{fig:nscaling_v1}
    \includegraphics[width=0.47\linewidth]{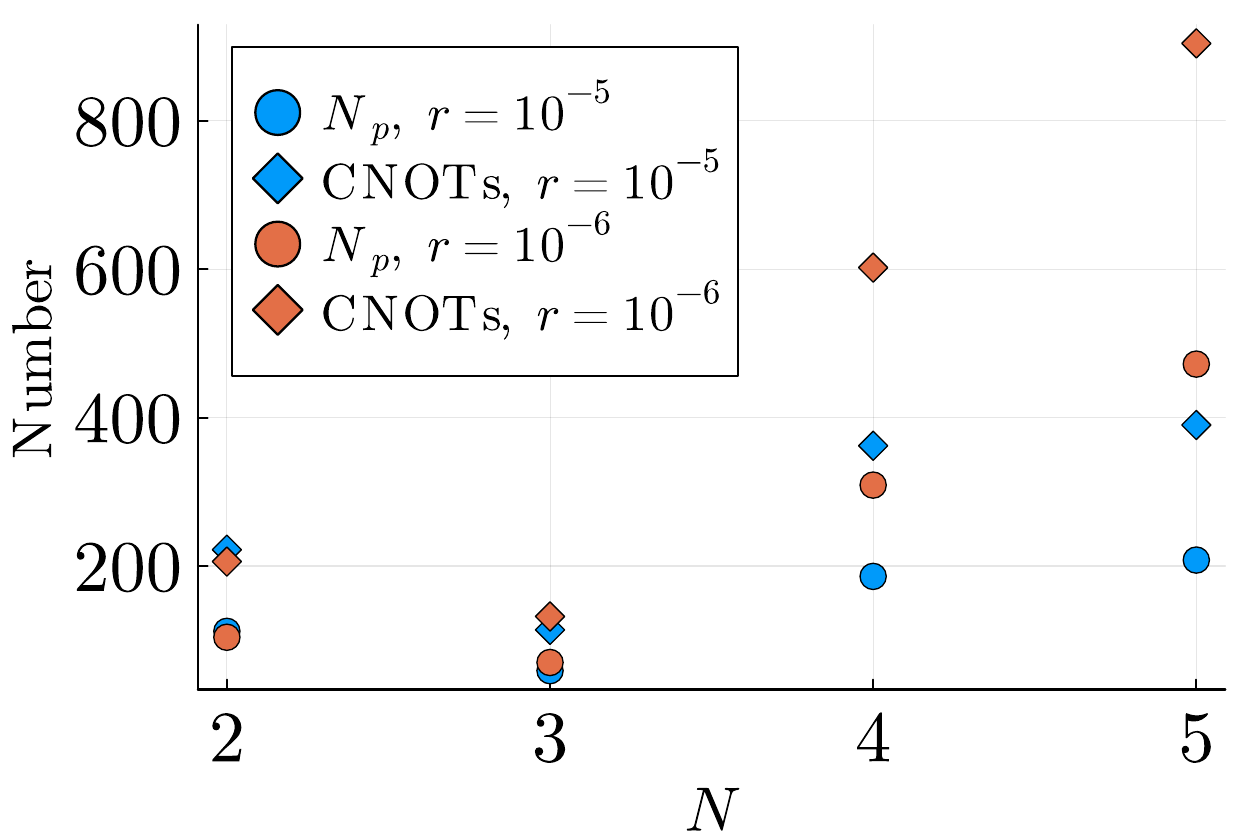}
    }%
    \\
    \subfloat[]{
    \label{fig:nscaling_t2}
    \includegraphics[width=0.47\linewidth]{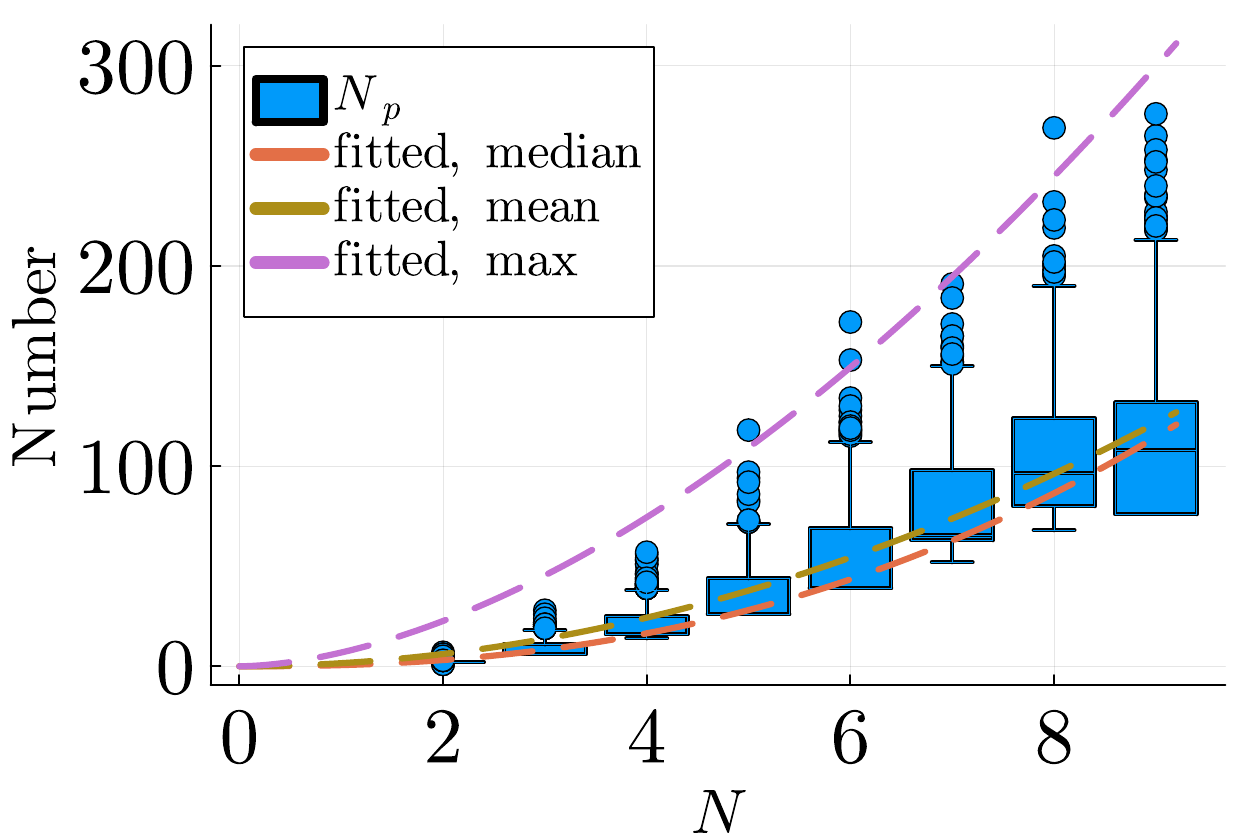}
    }
    \hfill
    \subfloat[]{%
    \label{fig:nscaling_v2}
    \includegraphics[width=0.47\linewidth]{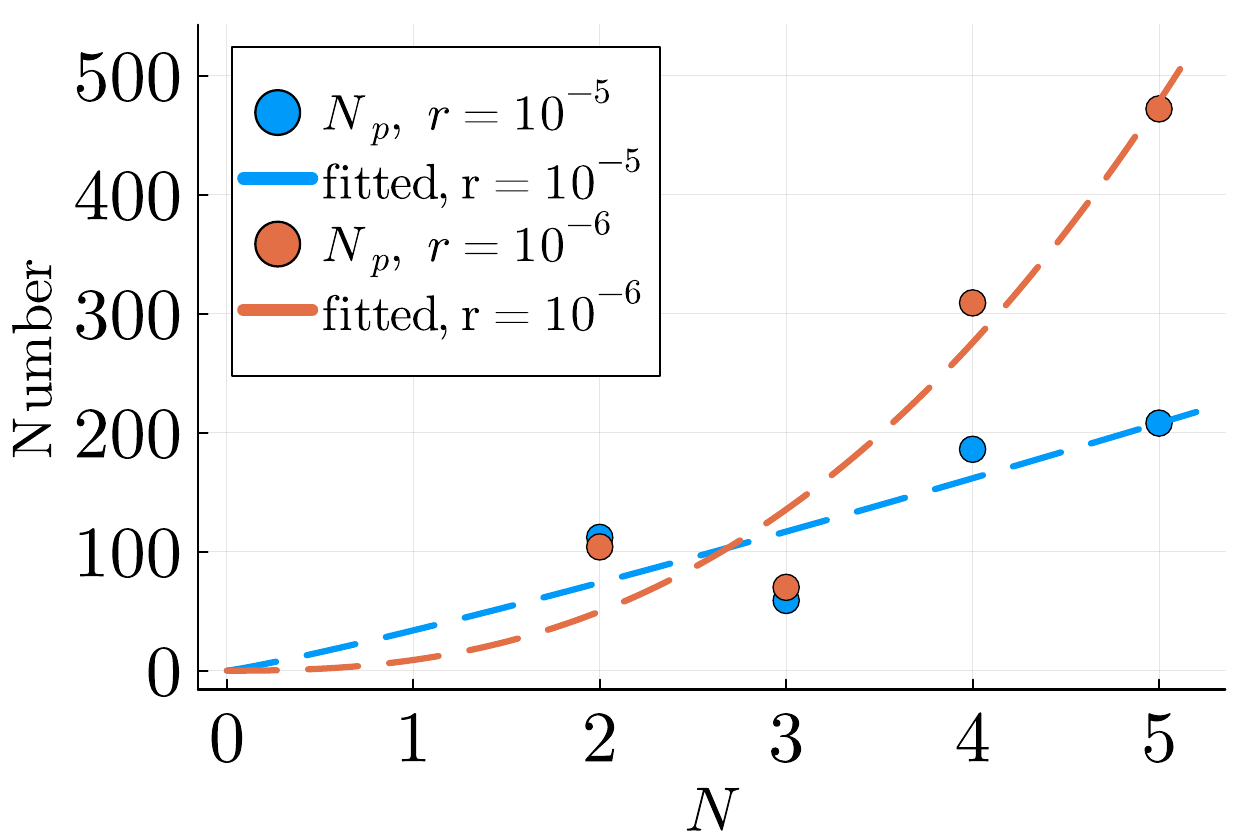}
    }%
    \caption{\textbf{Resource requirements for different system sizes.} (a) Ansatz and estimated CNOT counts for the trajectory method. (b) Ansatz and estimated CNOT counts for the vectorization method. (c) Ansatz counts fitted to a model of $aN^{b}$ for the trajectory method, with box plots of 1000 trajectories. (d) Ansatz counts fitted to a model of $aN^{b}$ for the vectorization method. In all the panels, $N_p$ represents the ansatz count, while $\mathrm{CNOTs}$ indicates the estimated CNOT count. The box plots in panels (a) and (c) are obtained from 1000 trajectories, and panels (b) and (d) display resource counts for two adaptive threshold choices. The results were obtained using the configuration described in Fig.~\ref{fig:4qubit_Z_vec}.
    }
    \label{fig:n4_scaling}
\end{figure*}

Finally, we discuss the resource requirement for our algorithms. Two key factors to consider are the total number of operators  and the number of multi-qubit operators in the ansatz, both of which are positively correlated with the final circuit depth required to execute the algorithm. 
To estimate the maximum number of CNOT gates needed for each time step, we start by calculating the number of CNOTs in the ansatz. The ansatz consists of unitaries that have the form $e^{-i\theta P_l}$, where $P_l$ is a Pauli word of length $l$.  Thus,  the total number of CNOT gates needed is given by  $\sum_{l>1}2(N_l - 1)$ where $N_l$ is the number of $P_l$s in the ansatz.
%Furthermore, the total number of CNOT gates can be estimated using twice the number of 2-qubit operators in the ans\"atze. 
%Due to the noise in NISQ quantum processors, the circuit depth is a major bottleneck for many quantum algorithms, which is why variational algorithms are preferable for near-term applications. 
The scaling of the ansatz and estimated CNOT counts versus the system size $N$ is shown in Figs.~\ref{fig:nscaling_t1} and~\ref{fig:nscaling_v1} for both the trajectory and vectorization methods. Please refer to Appendix~\ref{app:growth_ansatz} for a demonstration on how the sizes of the ans\"atze grow as operators are adaptively added. The largest CNOT count in the figure is around $900$ which is still far below the required CNOT count for any Trotterized algorithm using the same step size (with $dt = 0.01$, at least $1000$ trotter step is needed).

%Additionally, 
In order to study scaling of parameters and resources, 
the operator counts are fitted to either a polynomial of the form $a N^{b}$ or an exponential of the form $\alpha e^{\beta N}$ using least squares algorithm, and the results are summarized in Table.~\ref{table:2}. To compensate for the non-monotonic behavior of operator counts in the vectorization method, we add $(0, 0)$ to the fitting data. Since the operator count profiles for different trajectories in the trajectory method vary, we compute fitting data based on the (a) mean, (b) median and (c) maximum value of the operator counts at each time step of the trajectories. Based on the coefficient of determination ($R^2$) for both the polynomial and exponential fittings, it is clear that a low degree polynomial better fits our data. However, more data points are needed if we want to rule out the exponential model with high confidence. The curves for polynomial fitting are displayed in Figs.~\ref{fig:nscaling_t2} and~\ref{fig:nscaling_v2}.  
Based on the scaling data for the finite system sizes we studied, we do not observe signs of exponential scaling.  
%The resource count did not show signs of exponential scaling with respect to either the system size or the accuracy of the final result in the small size systems we studied. 
AVQDS has previously shown polynomial scaling in terms of the number of parameters and CNOTs for other spin models~\cite{Yao2021-is}. Confirming this observation for larger system in case of open quantum systems will be a topic for future studies.

\begin{table}[!ht]
\centering
\resizebox{\columnwidth}{!}{%
\begin{tabular}{||c | c | c | c ||} 
 \hline
\diagbox{Data}{Fitting} & $b$ & $R^2$ ($\mathrm{poly}$) & $R^2$ ($\mathrm{exp}$)  \\  
 \hline\hline
 Trajectory (median) & $2.39$ & $0.986$ & $0.961$ \\
 Trajectory (mean) & $1.98$ & $0.980$ & $0.938$ \\
 Trajectory (max) & $1.72$ & $0.972$ & $0.920$ \\
 Vectorization ($r=10^{-5}$) & $1.13$ & $0.821$ & $0.795$ \\
Vectorization ($r=10^{-6}$) & $2.47$ & $0.945$ & $0.943$\\
 \hline
\end{tabular}%
}
\caption{\textbf{A summary of fitting results.} The ansatz counts shown in Fig.~\ref{fig:n4_scaling} are fitted to either a polynomial of the form $a N^{b}$ or an exponential of the form $\alpha e^{\beta N}$ using a least squares algorithm. The table reports the coefficient of determination ($R^2$) for both fittings and the exponent of the resulting polynomial.}
\label{table:2}
\end{table}

\subsection{Hardware results}
\label{sec:hardware}
To showcase our algorithm on a quantum computer that is currently available, we save the parameters of the time evolution that are computed classically using UAVQDS, and use these parameters to measure the energy at each time step by executing the circuits for Eq.~\eqref{eq:hamit_t} on a real quantum computer. We run our experiments on IBM's 27-qubit processor \texttt{ibmq\_kolkata} with the Falcon architecture. To ensure the algorithms run smoothly, it is necessary to carefully compile the circuit based on the specific quantum device being used and to apply error suppression and mitigation strategies to obtain accurate and dependable results.

We note that we only simulate the dephasing model on the hardware because we want to avoid the additional complexity of mid-circuit measurement or reset operations required by simulating the amplitude damping model using the trajectory method (see Appendix~\ref{app:jumps}). Additionally, we compare the vectorization and trajectory method by selecting their respective $N$s such that they both use $4$ physical qubits, meaning $N=2$ for the trajectory method and $N=4$ for the vectorization method.

\textit{Circuit generation--}
We utilize the Qiskit transpiler to generate multiple circuits with varying numbers of CNOTs due to the stochastic addition of swap gates, and select the circuit with the lowest number of CNOTs for our study. Direct measurement is used for the vectorization method, while indirect measurement via the Hadamard test is used for the trajectory method (see Appendix~\ref{app:mO}). To synthesize the unitary required for measurements in the vectorization method, we use the Berkeley Quantum Synthesis Toolkit (BQSKit)~\cite{Younis2021-qs, szasz2023numerical}. 
% It is worth noting that, at the time of this paper, the function to synthesize an arbitrary state preparation unitary was only available in the \textit{1.1-dev} branch of BQSKit. 
We do not perform additional recompilation besides the optimization supported by the Qiskit transpiler.

\textit{Error mitigation and post processing--} 
We use the matrix-free measurement mitigation (M3) package to apply readout error mitigation in our study. This approach operates within a reduced subspace that is defined by the noisy input bitstrings requiring correction, making it scalable. The M3 package is described in references by~\cite{nation2021} and is accessible through Qiskit. We also incorporate dynamical decoupling using Qiskit's standard tools, implementing periodic gate sequences to suppress cross-talk and system-environment coupling~\cite{ezzell2022, Tripathi2022-ju,Pokharel2018-kh, Viola1999-ra}. Additionally, we apply an empirical error mitigation scheme based on the classical resolution enhancement (RE) technique, which has shown effectiveness in a different context~\cite{gomes2023computing}, to the trajectory simulations. In order to avoid the ambiguity of the value of enhancement factor, the RE method uses a convergence criteria based on tracing over the measured probabilities of the non-ancillary qubits in a Hadamard test setting. Since the vectorization method uses direct measurement without any ancillary qubit, we were unable to apply RE in this case. %However, we were unable to apply this technique to the vectorization method because we used direct measurement instead of the Hadamard test in this case. 
%\textbf{The previous sentence will require and explanation. People will not understand the what and why. NG: responded}. 
It is important to note that all of the error mitigation strategies we applied are out-of-the-box solutions and are resource-efficient, making them suitable for real-world applications.

\begin{figure*}[t]
    \centering
    \subfloat[]{
    \label{fig:6a}
    \includegraphics[width=0.48\linewidth]{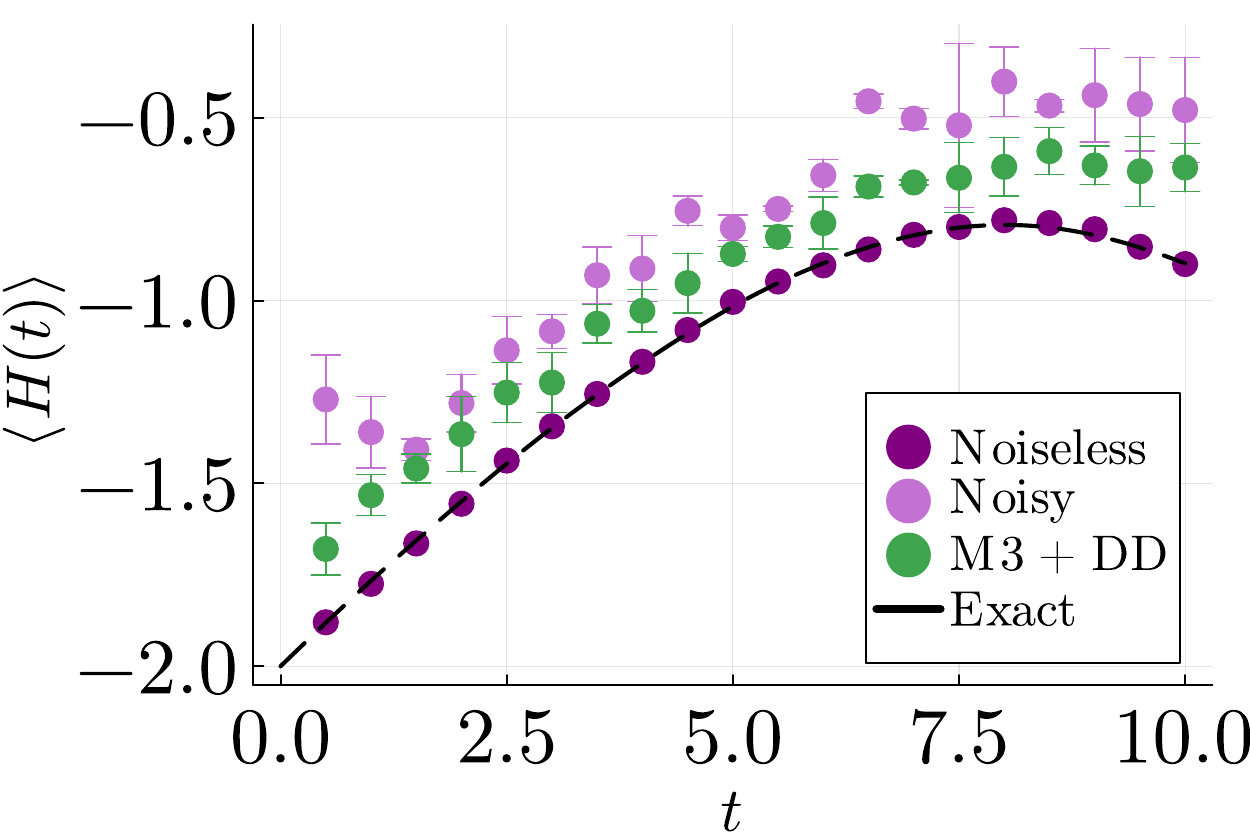}
    }
    \hfill
    \subfloat[]{%
    \label{fig:6b}
    \includegraphics[width=0.48\linewidth]{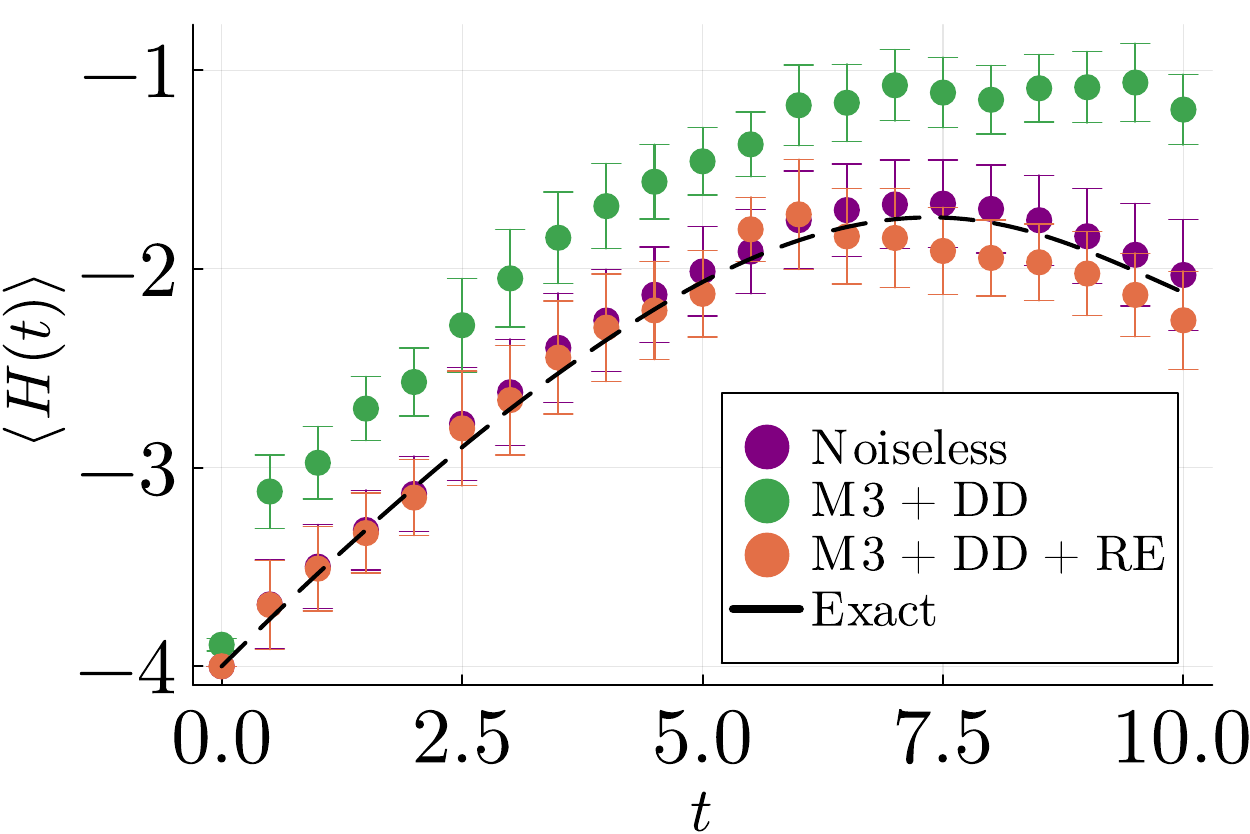}
    }%
    \caption{\textbf{Energy expectation value obtained by  executing the parameterized circuits on \texttt{ibmq\_kolkata}.} Panels (a) and (b) show the results for the vectorization method ($N=2$) and trajectory method ($N=4$), respectively. The noiseless, noisy, and exact curves represent the results from ideal circuit simulation, quantum computer without any error suppression or mitigation techniques and exact solver, respectively. Other curves show the results obtained from the quantum computer with corresponding error suppression or mitigation techniques: measurement error mitigation ($\mathrm{M3}$), dynamical decoupling (DD) and resolution enhancement (RE). The error bars in panel (a) represent two times the standard deviation from $3$ runs, and the error bars in panel (b) represent two times the standard error of the mean from $17$ trajectories. All curves from IBM quantum computers are obtained with $100000$ shots. Note that we do not have a noisy curve for the trajectory method because running trajectories on the real quantum computer is resource-intensive.}
    \label{fig:hardware}
\end{figure*}

\textit{Results--}
We show the experimental results on the hardware in Fig.~\ref{fig:hardware}. The figure shows $\expval{H}$ versus evolution time $t$ for vectorization (a) and trajectory (b) methods, respectively. It includes results obtained from ideal circuit simulation, quantum computer with and without error suppression or mitigation techniques, as well as the exact solver. The vectorization result is for a problem size of $N=2$, i.e, $4$ physical qubits. The trajectory method is for a problem size $N=4$, which requires $4$ physical qubits plus one additional ancilla qubit in order to measure $\expval{H}$ with the Hadamard test.  
%We observe qualitative agreement with the exact solution in the case of the vectorization method and quantitative agreement with the exact solution in the case of trajectory method. Even though the $N=2$ vectorization problem and $N=4$ trajectory problem requires similar numbers of ans\"atze (see Fig.~\ref{fig:n4_scaling}), the larger error of the vectorization algorithm is attributed to the fact that its ans\"atze includes non-local 2-qubit Pauli operators, which require swap gates when mapped to the actual hardware topology and lead to larger error in the final results. We expect the results to improve if the circuits are aggressively recompiled. On the other hand, 

From Figs.~\ref{fig:6a} and~\ref{fig:6b}
we observe good qualitative agreement between the exact solution and the results obtained from both the vectorization and trajectory methods. It is important to note that the error bars for the vectorization method and trajectory method represent distinct quantities. Specifically, the error bars for the vectorization method represent two times the standard deviation of $3$ runs, while the error bars for the trajectory method represent two times the standard error of the mean over 17 trajectories.   The performance of the vectorization method is particularly impressive because one would expect it to perform worse than the trajectory method, given that
%how little fine-tuning we performed. 
the $N=2$ vectorization problem requires more operators in the ansatz (see Fig.~\ref{fig:n4_scaling}) than the $N=4$ trajectory problem and its ans\"atze include non-local 2-qubit Pauli operators. In addition, we expect the results to improve significantly with aggressive circuit recompilation since minimal fine-tuning was done in our hardware runs. The results from the trajectory method improve considerably after post-processing with the RE technique. %\HC{Currently RE does not work with the vectorization method, should we really add it here?}
However, it is important to note that we have only  explored the RE method for a few limited cases and its generalizability as a potential error mitigation technique is still not fully understood. %Despite this, we report the RE results here because it is an interesting topic for future studies to understand where the improvement comes from and to explore the applicability of this method in a wider range of scenarios.
%However, there are a few caveats we would like to point out.
%The results from the trajectory method are surprisingly impressive, considering how little fine-tuning we performed. However, there are a few caveats we would like to point out. First, we only ran $17$ trajectories due to our limited access to IBM quantum computers. The error bar is still too large to make any statements beyond an accuracy of $\sim 0.1$. Second, our simulation parameters are chosen such that there are only on average $0.4$ jumps within the total evolution time. We expect the error to become larger for larger Lindbladian rates and longer evolution times. Third, the ansatz parameters of the final circuits are pre-generated classically, and the error from each time step would accumulate when running the entire process on real quantum computers. 
Summarizing, we see that the results from the IBM quantum computers are highly encouraging and suggest that simulating Lindblad equations is within the capability of NISQ devices.

\section{Conclusion}
\label{sec:conclusion}
We have introduced an adaptive variational quantum algorithm to simulate open quantum system dynamics described by a Lindblad equation. The method variationally solves either a vectorized or a stochastically unravelled Lindblad equation , with the ansatz built adaptively at every time step by minimizing McLachlan's distance in a greedy manner. We provide a benchmark of the algorithm's performance on both the ideal simulator and IBM's quantum processor for simulating the open-system quantum annealing dynamics, achieving good quantitative and qualitative agreement with the exact solution. Additionally, we perform a resource analysis on finite systems and find polynomial scaling, indicating the algorithm's potential to be extended to larger systems. Based on our numerical findings, we conjecture that both algorithms will exhibit polynomial scaling in gate complexity, given the appropriate selection of an ansatz pool. Additionally, these algorithms demonstrate a space-time trade-off (number of trajectories versus number of qubits) that mirrors what we observed in their classical counterparts. When considering NISQ-era quantum devices, the trajectory method is more favorable because of two reasons. First, qubits are more valuable as a resource in the NISQ era. Secondly, the vectorization method presents an additional challenge due to its non-local effective Hamiltonian, which requires more complex ans\"atze. 

Scaling our algorithm to sizes where meaningful simulations~\cite{khasseh2023active, Mattioni2021-av, Weimer2021-ys,Yao2021-nx} can be conducted on NISQ hardware presents several challenges, with the foremost being the required circuit depth. Despite considerable advancements in enhancing the quality of quantum processors, decoherence still sets a practical limit on the achievable circuit depth. For instance, the most advanced experiments on IBM's 127-qubit chip have been constrained to circuit depths of less than 100~\cite{Kim2023-qs}. In contrast, our algorithm's circuit depth surpasses this threshold even for problems involving just a few qubits, as illustrated in Fig.~\ref{fig:n4_scaling}, where the circuit depth closely follows the CNOT count. Besides waiting for better hardware, enhancements in both algorithmic design and software optimization are essential. For example, employing advanced recompilation techniques with heuristic algorithms can substantially reduce circuit depth~\cite{Younis2021-qs}; implementing advanced error mitigation strategies can allow for longer circuit~\cite{Urbanek2021-vz,Van_den_Berg2023-pe}. Additionally, the training part of our algorithm, currently performed classically, will ultimately need to be transitioned to a quantum computer. This transition will benefit from identifying the MCP to reducing the measurement overhead of the training process and developing strategies that ensures convergence of the training~\cite{Shkolnikov2023-kg}. Our algorithm pushes the boundary of NISQ algorithms for open quantum system simulation, providing a promising avenue for future research in this field.

%\textit{Discussion.---} 
%\section{Outlook}

\acknowledgments
The authors would like to thank Aaron Szasz for insightful discussions on BQSKit. This work was supported by the U.S. Department of Energy, Office of Science, National Quantum Information Science Research Centers, Quantum Systems Accelerator (HC, WAdJ) and by DOE Office of Advanced Scientific Computing Research (ASCR) through the ARQC program (NG, SN).

This research used resources of the Oak Ridge Leadership Computing Facility, which is a DOE Office of Science User Facility supported under Contract DE-AC05-00OR22725 and resources of the National Energy Research Scientific Computing Center (NERSC), a U.S. Department of Energy Office of Science User Facility located at Lawrence Berkeley National Laboratory.

The authors also acknowledge the use of IBM Quantum services. The views expressed are those
of the authors and do not reflect the official policy or position of IBM or the IBM Quantum team.

The supplementary code for this paper is available in the this GitHub \href{https://github.com/neversakura/Adaptive-Open-Quantum-System-Simulation}{repository}.

%\bibliography{sample.bib, refs.bib}
%\bibliography{refs.bib}
\bibliographystyle{quantum}
\bibliography{refs.bib}

%\onecolumn
\onecolumngrid
\newpage
\appendix
\counterwithin{figure}{section}
\counterwithin{table}{section}
\section{Frequently used symbols}
For convenience, we provide a table summarizing frequently used symbols in the main text.
\label{app:symbols}
\begin{table}[!ht]
\centering
\resizebox{\columnwidth}{!}{%
\begin{tabular}{||c | l ||} 
 \hline
 Symbol & Definition \\  
 \hline\hline
 $H(t)$ & Hamiltonian of the system of interest \\
 $\rho\pqty{t}$ & Density matrix of the system of interest \\
 $\mathcal{L}$ & Lindblad superoperator of the Lindblad equation \\
 $L_k$ & $k$th Lindblad operator of the Lindblad equation, with the dissipation rate being absorbed \\
 $\gamma_k$ & Dissipation rate associated with the $k$th Lindblad operator \\
 $\kett{\rho\pqty{t}}$ or $\mathrm{vec}(\rho(t))$ & Vectorized density matrix \\
 $H_\mathrm{vec}$  & Effective Hamiltonian for the vectorized Lindblad equation  \\
 $H_\mathrm{urv}$ & Effective Hamiltonian for the unravelled Lindblad equation \\
 $\bar{L}$ & Complex conjugate of operator $L$ \\
$H_\mathrm{eff}$ & Generic effective non-Hermitian Hamiltonian \\
$H_\mathrm{e}$ & Hermitian part of the effective Hamiltonian $H_\mathrm{eff}$\\
$H_\mathrm{a}$ & $-iH_\mathrm{a}$ is the anti-Hermitian part of the effective Hamiltonian $H_\mathrm{eff}$\\
$\mathcal{D}$ & Square of the McLachlan distance used in the variational principle\\
$\ket{\tilde{\psi}\pqty{t}}$ & Unnormalized state vector solution for the effective Schr\"odinger equation with $H_\mathrm{eff}$ \\
$\ket{\psi\pqty{t}}$ & Normalized state vector solution for the effective Schr\"odinger equation $H_\mathrm{eff}$\\
$\ket{\tilde{\psi}_j\pqty{t}}$ & Unnormalized state vector solution for the $j$th trajectory \\
$\ket{\psi_j\pqty{t}}$ & Normalized state vector solution for the $j$th trajectory \\
$ \ket{\phi\pqty{t}}$ & State from parameterized quantum circuits\\
$\ket{\psi_\mathrm{R}}$ & Reference state for the parameterized quantum circuits\\
$\theta_\mu\pqty{t}$ & $\mu$th ansatz parameter for the parameterized circuit \\
$A_\mu$ & $\mu$th ansatz operator for the parameterized circuit \\
$\bth\pqty{t}$ & Vector of the ansatz parameters $\bqty{\theta_1\pqty{t}, \theta_2\pqty{t}, \cdots \theta_k\pqty{t}}$\\
$\mathbf{A}$ & Vector of the ansatz operators $\bqty{A_1, A_2, \cdots A_k}$ \\
$\bM, \bV$ & Coefficient matrix and vector of the linear equation generated by the variational principle\\
$\Gamma$ & A classical register to record the norm of $\ket{\tilde{\psi}\pqty{t}}$ in the algorithm \\
$\ket{O^\dagger}$ & Normalized and vectorized operator $O^\dagger$\\ 
$X_i, Y_i, Z_i$ & Pauli $X$, $Y$ and $Z$ operators acting on the $i$th qubit \\
$P_i$ & A Pauli $X$, $Y$ or $Z$ matrix acting on the $i$th qubit \\
$\mathcal{P}_{s}$ & Notation for different operator pools\\ 
$L_i^+, L_i^-$ & Plus and minus Lindblad operators acting on the $i$th qubit \\
$\gamma_i^+, \gamma_i^-$ & Dissipation rates associated with the plus and minus Lindblad operators \\
$A\pqty{t}, B\pqty{t}$ & Annealing schedules (scalar functions)\\
$H_\mathrm{D}, H_\mathrm{P}$ & Driving and problem Hamiltonian in a standard quantum annealing protocol \\
 $N_p$ & Total number of parameter in the ans\"atze at the end of evolution\\
$D $ & Trace-norm distance between
the solution of variation algorithm and the exact solution \\
 \hline
\end{tabular}%
}
\caption{\textbf{List of symbols and their definitions used in the main text.}}
\label{table:symbols}
\end{table}
\section{Derivation of the vectorized Lindblad equation}
\label{appendix:vectorized_lindblad}
In this section we show how to derive the vectorized Lindblad equation from
\begin{equation}
    \label{app:lind_eq}
    \frac{d}{dt}\rho\pqty{t} = -i\comm{H\pqty{t}}{\rho\pqty{t}}+\Lag\bqty{\rho\pqty{t}} \ ,
\end{equation}
where the dissipative term $\Lag\bqty{\rho\pqty{t}}$ is
\begin{equation}
    \Lag\bqty{\rho\pqty{t}} = \sum_{k=1}^{K}  \pqty{L_k \rho\pqty{t} L^\dagger_k -\frac{1}{2}\acomm{L^\dagger_k L_k}{\rho\pqty{t}}} \ .
\end{equation}
The strategy is to apply the identity $\operatorname{vec}(A B C)=(C^{\mathrm{T}} \otimes A)\operatorname{vec}(B)$ to each term on the right-hand side of Eq.~\eqref{app:lind_eq}. The Hamiltonian part $\comm{H\pqty{t}}{\rho\pqty{t}}$ can be written in terms of

\begin{equation}
    \label{app:Hvec}
    \operatorname{vec} \Bqty{  \rho\pqty{t} H\pqty{t}} = \pqty{H^\mathrm{T}\pqty{t}\otimes I}\operatorname{vec}\pqty{\rho\pqty{t}}, \quad
    \operatorname{vec} \Bqty{ H\pqty{t} \rho\pqty{t}} = \pqty{I\otimes H\pqty{t}}\operatorname{vec}\pqty{\rho\pqty{t}} \ .
\end{equation}
The Lindblad part can be written in terms of
\begin{subequations}
\label{app:Lvec}
\begin{gather}
    \operatorname{vec} \pqty{L_k \rho L^\dagger_k} = \pqty{\bar{L}_k\otimes L_k }\operatorname{vec}\pqty{\rho} ,\quad \operatorname{vec} \pqty{L^\dagger_k L_k \rho } = \pqty{I \otimes L^\dagger_k L_k }\operatorname{vec}\pqty{\rho} , \\
    \operatorname{vec}\pqty{ \rho L^\dagger_k L_k} =  \pqty{L^\mathrm{T}_k \bar{L}_k \otimes I} \operatorname{vec}\pqty{\rho} \ .
\end{gather}
\end{subequations}
Summing up every term in Eqs.\eqref{app:Hvec} and~\eqref{app:Lvec}, the effective Hamiltonian can be obtained as
\begin{equation}\label{app:Heff}
    H_\mathrm{vec} = I\otimes H - H^T\otimes I 
    +i\sum_k \bqty{\bar{L}_k\otimes L_k-\frac{1}{2}\pqty{I\otimes L_k^{\dagger} L_k + L_k^{T} \bar{L}_k\otimes I}} \ .
\end{equation}

\section{Vectorization in the Pauli basis}\label{appendix:vectorization_pauli_basis}
An alternative approach to vectorizing the ME involves expressing the density matrix in the basis of Pauli matrices and representing the superoperators through the Pauli transfer matrices (PTMs). However, this method leads to a notable increase in circuit complexity for two primary reasons. 
\begin{itemize}
    \item Vectorization can transform a single-qubit pure state into a more complex state, such as a 2-qubit Bell state. For instance, consider the state $\dyad{0}$. In the Pauli basis arranged in lexicographic order, this state, when vectorized, would be represented as  $(\ket{00}+\ket{11})/\sqrt{2}$ (up to a constant), necessitating an additional layer of 2-qubit gates for state preparation.
    \item When employing the column-stacking method for vectorization, if every term in the Hamiltonian is 2-local and each Lindblad operator is 1-local, then each term in the resultant effective Hamiltonian (as per Eq.~\eqref{app:Heff}) remains at most 2-local. For example, the PTM for the superoperator $\comm{ZZ}{\cdot}$ is $(XXXY + XYXX+YXYY+YYYX-XXYX-XYYY-YXXX-YYXY)/2$. To implement the Hamiltonian pool~\cite{Yao2021-is} would require  an arbitrary angle 4-qubit Pauli rotation, substantially increasing the number of required CNOT gates.
\end{itemize}

Given these considerations, particularly the worse error rates associated with 2-qubit gates on contemporary quantum devices, we opted against this vectorization approach to reduce the circuit depth. The potential for a more efficient implementation of this approach remains an intriguing area for future research.

\section{Derivation of the evolution equation}\label{appendix:variational_alg}

In this appendix, we derive the evolution equation of the variational parameters for the effective Schr\"odinger equation
\be
    \label{app_eq:effective_schordinger}
    \ket*{\dot{\psi}} = -i H_\mathrm{eff}\ket{\psi} \ ,
\ee
where the dot symbol denotes the time derivative. Before proceeding, it is worth mentioning that the effective Hamiltonian in Eq.~\eqref{app_eq:effective_schordinger} is not necessarily Hermitian. The ansatz state is parameterized by $\bth\pqty{t}\equiv\bqty{\theta_1\pqty{t}, \theta_2\pqty{t}, \cdots \theta_k\pqty{t}}^T$
\begin{equation}
    \label{app_eq:ansatz}
    \ket{\phi\pqty{\bth\pqty{t}}} = \prod_{\mu=1}^{k} e^{-i\theta_\mu\pqty{t} A_\mu} \ket{\psi_\mathrm{R}} \ ,
\end{equation}
where $A_\mu$ and $\ket{\psi_\mathrm{R}}$ are the ansatz operators and the reference state, respectively.
It is more convenient to derive the evolution equation of $\bth\pqty{t}$ using the density matrix representation to avoid the problem of a time dependent global phase~\cite{Yuan2019-ei}. Let $W\pqty{\bth}\equiv\dyad{\phi\pqty{\bth}}{\phi\pqty{\bth}}$, the equation of motion is given by
\begin{subequations}
\label{app_eq:W_eom}
\begin{align}
    \dot{W} &= \dyad*{\dot{\phi}}{\phi}+\dyad*{\phi}{\dot{\phi}} \label{app_eq:Wdot}\\
    &= -i\pqty{\Heff W - W \Heff^\dagger} \label{app_eq:Wcomm}\ ,
\end{align}
\end{subequations}
where we make use of Eq.~\eqref{app_eq:effective_schordinger}
in going from Eq.~\eqref{app_eq:Wdot} to Eq.~\eqref{app_eq:Wcomm}. Applying the McLachlan’s variational principle to Eq.~\eqref{app_eq:W_eom}, we have
\be
\label{app_eq:W_var_prin}
\delta \norm{\frac{dW\pqty{\bth\pqty{t}}}{dt}-\Lag\bqty{W\pqty{\bth\pqty{t}}}}^{2} =0 \ ,
\ee
where $\Lag\bqty{W}\equiv  -i\pqty{\Heff W - W \Heff^\dagger}$.
Let $\mathcal{D}\equiv\norm{\frac{dW\pqty{\bth\pqty{t}}}{dt}-\Lag\bqty{W\pqty{\bth\pqty{t}}}}^{2}$, the McLachlan distance (which is actually $\sqrt{\mathcal{D}}$) can be simplified as
\be
    \label{app_eq:D_terms}
    \mathcal{D}=\Tr\bqty{\pqty{\dot{W}^\dagger -\Lag^\dagger\bqty{W}}\pqty{\dot{W}-\Lag\bqty{W}}} = \Tr\bqty{\dot{W}^\dagger \dot{W}} -2\mathrm{Re}\Bqty{\Tr\Bqty{\dot{W}^\dagger\Lag\bqty{W}}}+\Tr\Bqty{\Lag^\dagger\bqty{W}\Lag\bqty{W}} \ .
\ee

Based on Eq.~\eqref{app_eq:Wdot}, $\dot{W}$ is Hermitian, i.e., $\dot{W}=\dot{W}^\dagger$. The first term $\Tr\bqty{\dot{W}^\dagger \dot{W}}$ on the RHS of Eq.~\eqref{app_eq:D_terms} can be simplified by the chain rule
\be
\Tr\Bqty{\dot{W}^\dagger \dot{W}} = \sum_{\mu,\nu}\Tr \Bqty{\pdv{W}{\theta_\mu}\pdv{W}{\theta_\nu}} \dot{\theta}_\mu \dot{\theta}_\nu = \dot{\bth}^T \bM \dot{\bth} \ ,
\ee
where $\bM$ is a matrix with elements
\begin{subequations}
\label{app_eq:M}
\begin{align}
    \bM_{\mu \nu} &=\Tr\Bqty{\pdv{W}{\theta_\mu}\pdv{W}{\theta_\nu}} \\ 
    &=\Tr\Bqty{\pqty{ \pdv{\ket{\phi}}{\theta_\mu}\bra{\phi} + \ket{\phi}\pdv{\bra{\phi}}{\theta_\mu} }\pqty{ \pdv{\ket{\phi}}{\theta_\nu}\bra{\phi} + \ket{\phi}\pdv{\bra{\phi}}{\theta_\nu} }}\\
    &= 2\mathrm{Re}\pqty{\pdv{\bra{\phi}}{\theta_{\mu}}\pdv{\ket{\phi}}{\theta_{\nu}} +\bra{\phi}\pdv{\ket{\phi}}{\theta_{\mu}} \bra{\phi}\pdv{\ket{\phi}}{\theta_{\nu}}} \ .
\end{align}
\end{subequations}
Because $\theta_\mu$ are real parameters, $\bM$ is a real symmetric matrix.

Similarly, the second term on the RHS of Eq.~\eqref{app_eq:D_terms} can be simplified as
\begin{subequations}
\begin{align}
    2\mathrm{Re}\Bqty{\Tr\Bqty{\dot{W}^\dagger\Lag\bqty{W}}} &= 2\mathrm{Re}\Bqty{\Tr\Bqty{\pdv{W}{\theta_\mu} \Lag\bqty{W}}}\dot{\theta}_\mu \\
    &=2 \bV^T \dot{\bth} \ ,
\end{align}
\end{subequations}
where $\bV$ is a vector with elements
\begin{subequations}
\begin{align}
    \bV_\mu &= \mathrm{Re}\Bqty{\Tr\Bqty{\pdv{W}{\theta_\mu}\Lag\bqty{W}}}\\
    &=\mathrm{Im}\Bqty{\Tr\Bqty{ \pqty{ \pdv{\ket{\phi}}{\theta_\mu}\bra{\phi} + \ket{\phi}\pdv{\bra{\phi}}{\theta_\mu}} \pqty{\Heff \dyad{\phi}{\phi}-\dyad{\phi}{\phi} \Heff^\dagger}}} \\
    &=2\mathrm{Im}\pqty{\expval{\Heff}\bra{\phi}\pdv{\ket{\phi}}{\theta_\mu}+\pdv{\bra{\phi}}{\theta_\mu}\Heff \ket{\phi}} \ ,
\end{align}
\end{subequations}
and $\expval{\Heff} = \mel{\phi}{\Heff}{\phi}$.

Before calculating the last term on the RHS of Eq.~\eqref{app_eq:D_terms}, we split the effective Hamiltonian into a Hermitian and an anti-Hermitian parts
\be
\label{app_eq:split_H}
H_\mathrm{eff} = \frac{H_\mathrm{eff}+H^\dagger_\mathrm{eff}}{2} -i\pqty{ i\frac{H_\mathrm{eff}-H^\dagger_\mathrm{eff}}{2}} \equiv H_\mathrm{e} - i H_\mathrm{a} \ .
\ee
Then the last term becomes
\begin{align}
    \Tr\Bqty{\Lag^\dagger\bqty{W}\Lag\bqty{W}} &= \Tr\Bqty{\pqty{W\Heff^\dagger-\Heff W}\pqty{\Heff W - W \Heff^\dagger}} \notag\\
    &=\Tr\bqty{W\Heff^\dagger\Heff W}-\Tr\bqty{W\Heff^\dagger W \Heff^\dagger} - \Tr\bqty{\Heff W \Heff W} + \Tr\bqty{\Heff W W \Heff^\dagger} \notag\\
    &=\expval{\pqty{\He + i \Ha}\pqty{\He -i\Ha}} - \expval{\He + i\Ha}^2 - \expval{\He - i\Ha}^2 + \expval{\pqty{\He+i\Ha}\pqty{\He-i\Ha}} \notag\\
    &=2\expval{\He^2} - 2\expval{\He}^2 + 2\expval{\Ha^2} + 2\expval{\Ha}^2 +2i\expval{\comm{\Ha}{\He}} \ .
\end{align}
%For simplicity, we also use the notation $C\equiv \Tr\Bqty{\Lag^\dagger\bqty{W}\Lag\bqty{W}}$. 
The variational principle (Eq.~\eqref{app_eq:W_var_prin}) then yields
\be
\delta{\mathcal{D}\pqty{\dot{\bth}}} = \dot{\bth}^T \pqty{\bM + \bM^T} - 2\bV^T = 2\bM \dot{\bth} - 2\bV =  0 \ .
\ee
Finally, the evolution equation of the variational parameters is given by a linear equation
\be
\bM \dot{\bth} = \bV \ .
\ee

\section{Measurement circuit for $\bM$ and $\bV$}
\label{app:MV_measurement}
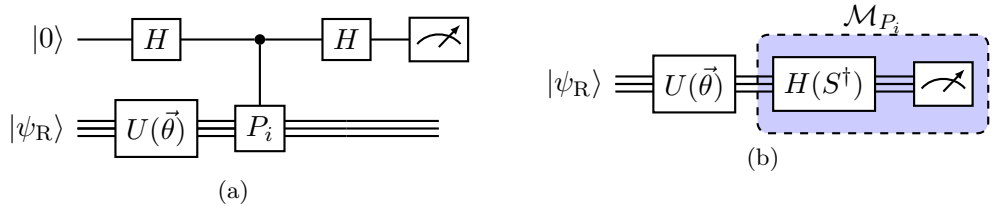
\begin{figure}[htb]
    \centering
    \subfloat[]{
    \begin{quantikz}
    \lstick{$\ket{0}$} & \gate{H} & \ctrl{1} & \gate{H} & \meter{} \\
        \lstick{$\ket{\psi_\mathrm{R}}$} &  \gate{U(\vec\theta)}\qwbundle[
    alternate]{} & \gate{P_i} \qwbundle[
    alternate]{} & \qwbundle[
    alternate]{} & \qwbundle[
    alternate]{}
    \end{quantikz}
    }%
    \qquad
    \subfloat[]{%
    \begin{quantikz}
        \lstick{$\ket{\psi_\mathrm{R}}$} &  \gate{U(\vec\theta)}\qwbundle[
        alternate]{} & \gate{H(S^\dagger)} \qwbundle[
        alternate]{}
        \gategroup[1,steps=2, style={dashed,
        rounded corners, fill=blue!20, inner xsep=2pt}, background]{{$\mathcal{M}_{P_i}$}}& \meter{}\qwbundle[
        alternate]{}
    \end{quantikz}
    }%
    \caption{Quantum circuit for evaluating $\ev{P_i}{\phi}$ using (a) Hadamard test; (b) direct measurements. $U\pqty{\vec\theta}$ denotes the unitary generated by the variational circuit $\prod_{\mu=1}^{k} e^{-i\theta_\mu\pqty{t} A_\mu}$. $H\pqty{S^\dagger}$ denotes the optional Hadamard (Hadmard-phase) gate to rotate the basis of $P_i$ to $z$ direction. The gates within the blue block implement a projective measurement onto the eigenbasis of $P_i$.} 
    \label{fig:measure_pauli}
\end{figure}

In this appendix, we will discuss the quantum circuit used to measure the parameter in the variational equation of motion $\bM \dot{\bth} = \bV$, where
\be
\label{app:M}
    \bM_{\mu \nu} =2\mathrm{Re}\pqty{\pdv{\bra{\phi}}{\theta_{\mu}}\pdv{\ket{\phi}}{\theta_{\nu}} +\bra{\phi}\pdv{\ket{\phi}}{\theta_{\mu}} \bra{\phi}\pdv{\ket{\phi}}{\theta_{\nu}}} \ ,
\ee
and 
\be
\label{app:V}
    \bV_\mu = 2\mathrm{Im}\pqty{\expval{\Heff}\bra{\phi}\pdv{\ket{\phi}}{\theta_\mu}+\pdv{\bra{\phi}}{\theta_\mu}\Heff \ket{\phi}} \ .
\ee
assuming that each ansatz operator $A_\mu$ is a Pauli operator. We only present the resulting circuits here and encourage interested readers to refer to~\cite{Mitarai2019-pv} for further details. First, the term $\expval{H_\mathrm{eff}}$ can be evaluated by measuring all the Pauli strings $P_i$ that make up the Hamiltonian. The quantum circuit for both direct and indirect measurement (Hadamard test) are shown in Fig.~\ref{fig:measure_pauli}, where we use notation $U\pqty{\vec\theta}$ to denote the the unitary generated by the variational circuit and $H\pqty{S^\dagger}$ to denote the optional Hadamard (Hadamard-phase) gate to rotate the basis of $P_i$ to $z$ direction.

Second, the term $\bra{\phi}\pdv{\ket{\phi}}{\theta_\mu}$ and $\bra{\phi}H_\mathrm{eff}\pdv{\ket{\phi}}{\theta_\mu}$ can be evaluated by using a generalized Hadamard test or the corresponding direct measurement circuit. The main observation is that, using
\begin{equation}
    \pdv{\ket{\phi}}{\theta_\mu} = -i \prod_{l=\mu+1}^{k} e^{-i\theta_l A_l}  A_\mu \prod_{j=1}^{\mu} e^{-i\theta_j A_j} \ket{\psi_\mathrm{R}} \ ,
\end{equation}
the term $\bra{\phi}P_i\pdv{\ket{\phi}}{\theta_\mu}$ can be simplified as
\begin{equation}
    \label{app:PA_measure}
    \bra{\phi}P_i\pdv{\ket{\phi}}{\theta_\mu} \propto \ev{U^\dagger_{\mu:1}U^\dagger_{k:\mu+1}P_i U_{k:\mu+1} A_\mu U_{\mu:1}}{\psi_\mathrm{R}} \ ,
\end{equation}
where $U_{\mu:1} = \prod_{j=1}^{\mu} e^{-i\theta_j A_j}$ and $U_{k:\mu+1} = \prod_{l=\mu+1}^{k} e^{-i\theta_l A_l}$. Eq.~\eqref{app:PA_measure} can be evaluated using the generalized Hadamard test (Fig.~\ref{fig:g_hadamard}) or the corresponding direct measure circuit (Figs.~\ref{fig:g_dm1} and~\ref{fig:g_dm2}).

\begin{figure}[htb]
    \centering
    \subfloat[]{
    \begin{quantikz}
    \lstick{$\ket{0}$} & \gate{H} & \ctrl{1} & \qw & \ctrl{1} & \gate{S^b} & \gate{H} & \meter{} \\
        \lstick{$\ket{\psi_\mathrm{R}}$} &  \gate{U_{\mu:1}}\qwbundle[
    alternate]{} & \gate{A_\mu} \qwbundle[
    alternate]{} & \gate{U_{k:\mu+1}}\qwbundle[
    alternate]{} &\gate{P_i}\qwbundle[
    alternate]{} &\qwbundle[
    alternate]{} &\qwbundle[
    alternate]{} &\qwbundle[
    alternate]{}
    \end{quantikz}
    \label{fig:g_hadamard}
    }%
    \\
    \subfloat[]{%
    \begin{quantikz}
        \lstick{$\ket{\psi_\mathrm{R}}$} &  \gate{U_{\mu:1}}\qwbundle[
        alternate]{} & \gate{H(S^\dagger)} \qwbundle[
        alternate]{}
        \gategroup[1,steps=2, style={dashed,
        rounded corners, fill=blue!20, inner xsep=2pt}, background]{{$\mathcal{M}_{A_\mu}$}}& \meter{}\qwbundle[
        alternate]{}& \gate{U_{k:\mu+1}}\qwbundle[
        alternate]{} & \gate{H(S^\dagger)} \qwbundle[
        alternate]{}
        \gategroup[1,steps=2, style={dashed,
        rounded corners, fill=blue!20, inner xsep=2pt}, background]{{$\mathcal{M}_{P_i}$}}& \meter{}\qwbundle[
        alternate]{}
    \end{quantikz}
    \label{fig:g_dm1}
    }%
    \\
    \subfloat[]{
    \begin{quantikz}
        \lstick{$\ket{\psi_\mathrm{R}}$} &  \gate{U_{\mu:1}}\qwbundle[
        alternate]{} & \gate{e^{\pm i \pi A_\mu / 4}} \qwbundle[
        alternate]{}
        & \gate{U_{k:\mu+1}}\qwbundle[
        alternate]{} & \gate{H(S^\dagger)} \qwbundle[
        alternate]{}
        \gategroup[1,steps=2, style={dashed,
        rounded corners, fill=blue!20, inner xsep=2pt}, background]{{$\mathcal{M}_{P_i}$}}& \meter{}\qwbundle[
        alternate]{}
    \end{quantikz}
    \label{fig:g_dm2}
    }%
    \caption{Quantum circuits for evaluating $\bra{\phi}P_i\pdv{\ket{\phi}}{\theta_\mu}$ using (a) generalized Hadamard test; (b) and (c) direct measurements. The generalized Hadamard test circuit uses a binary integer $b \in \{0,1\}$ as an input. If $b=0$, the circuit measures the real part of $\ev{U^\dagger_{\mu:1}U^\dagger_{k:\mu+1}P_i U_{k:\mu+1} A_\mu U_{\mu:1}}{\psi_\mathrm{R}}$, and if $b=1$, it measures the imaginary part. For direct measurement, circuit (b) measures the real part of the targeted quantity, while circuit (c) measures the imaginary part.}
    \label{fig:measure_gpauli}
\end{figure}

Finally, the metric tensor, given by $\pdv{\bra{\phi}}{\theta_{\mu}}\pdv{\ket{\phi}}{\theta_{\nu}}$, can also be evaluated using circuits similar to those discussed above. We note that this expression can be simplified as
\begin{equation}
    \pdv{\bra{\phi}}{\theta_{\mu}}\pdv{\ket{\phi}}{\theta_{\nu}} \propto \ev{U^\dagger_{\mu:1}A_\mu U_{\mu+1:\nu+1} A_\nu U_{\nu:1}}{\psi_\mathrm{R}} \ ,
\end{equation}
assuming $\mu \ge \nu+1$. Quantum circuits for evaluating this quantity are shown in Fig.~\ref{fig:measure_metric_tensor}.
\begin{figure}[htb]
    \centering
    \subfloat[]{
    \begin{quantikz}
    \lstick{$\ket{0}$} & \gate{H} & \ctrl{1} & \gate{X}\qw & \ctrl{1} & \gate{X}\qw & \gate{S^b} & \gate{H} & \meter{} \\
        \lstick{$\ket{\psi_\mathrm{R}}$} &  \gate{U_{\nu:1}}\qwbundle[
    alternate]{} & \gate{A_\nu} \qwbundle[
    alternate]{} & \gate{U_{\mu:\nu+1}}\qwbundle[
    alternate]{} &\gate{A_\mu}\qwbundle[
    alternate]{} &\qwbundle[
    alternate]{} &\qwbundle[
    alternate]{} &\qwbundle[
    alternate]{} &\qwbundle[
    alternate]{}
    \end{quantikz}
    }%
    \\
    \subfloat[]{%
    \begin{quantikz}
        \lstick{$\ket{\psi_\mathrm{R}}$} &  \gate{U_{\nu:1}}\qwbundle[
        alternate]{} & \gate{H(S^\dagger)} \qwbundle[
        alternate]{}
        \gategroup[1,steps=2, style={dashed,
        rounded corners, fill=blue!20, inner xsep=2pt}, background]{{$\mathcal{M}_{A_\nu}$}}& \meter{}\qwbundle[
        alternate]{}& \gate{U_{\mu:\nu+1}}\qwbundle[
        alternate]{} & \gate{H(S^\dagger)} \qwbundle[
        alternate]{}
        \gategroup[1,steps=2, style={dashed,
        rounded corners, fill=blue!20, inner xsep=2pt}, background]{{$\mathcal{M}_{A_\mu}$}}& \meter{}\qwbundle[
        alternate]{}
    \end{quantikz}
    }%
    \\
    \subfloat[]{
    \begin{quantikz}
        \lstick{$\ket{\psi_\mathrm{R}}$} &  \gate{U_{\nu:1}}\qwbundle[
        alternate]{} & \gate{e^{\pm i \pi A_\nu / 4}} \qwbundle[
        alternate]{}
        & \gate{U_{\mu:\nu+1}}\qwbundle[
        alternate]{} & \gate{H(S^\dagger)} \qwbundle[
        alternate]{}
        \gategroup[1,steps=2, style={dashed,
        rounded corners, fill=blue!20, inner xsep=2pt}, background]{{$\mathcal{M}_{A_\mu}$}}& \meter{}\qwbundle[
        alternate]{}
    \end{quantikz}
    }%
    \caption{Quantum circuits for evaluating $\pdv{\bra{\phi}}{\theta_{\mu}}\pdv{\ket{\phi}}{\theta_{\nu}}$ using (a) generalized Hadamard test; (b) and (c) direct measurements. The generalized Hadamard test circuit uses a binary integer $b \in \{0,1\}$ as an input. If $b=0$, the circuit measures the real part of $\ev{U^\dagger_{\mu:1}A_\mu U_{\mu+1:\nu+1} A_\nu U_{\nu:1}}{\psi_\mathrm{R}}$ (for $\mu \ge \nu$), and if $b=1$, it measures the imaginary part. For direct measurement, circuit (b) measures the real part of the targeted quantity, while circuit (c) measures the imaginary part.}
    \label{fig:measure_metric_tensor}
\end{figure}

\section{Evolution of the state-vector norm}
\label{app:norm_evo}
In this section, we show how to track the evolution of the state-vector norm by measuring the anti-Hermitian component of the effective Hamiltonian at every time step. We start with the Schr\"odinger equation
\begin{equation}
    \frac{d}{dt}\ket{\tilde{\psi}\pqty{t}} = -i H_\mathrm{eff}\ket{\tilde{\psi}\pqty{t}} \ ,
\end{equation}
with $\ket{\tilde{\psi}\pqty{t}}$ being the unnormalized state vector and $\ket{\psi\pqty{t}}$ being the normalized state vector. The effective Hamiltonian has both the Hermitian $H_\mathrm{e}$ and anti-Hermitian parts $-iH_\mathrm{a}$, i.e.,
$
    H_\mathrm{eff} = H_\mathrm{e} - i H_\mathrm{a}
$. We also use the notation $\ket{\phi\pqty{t}}$ for the state generated by our variational circuit. The evolution of $\braket{\tilde{\psi}\pqty{t}}$ can be calculated as
\begin{subequations}
\begin{align}
    \frac{d\braket{\tilde{\psi}\pqty{t}}}{dt} &= \frac{d\bra{\tilde\psi \pqty{t}}}{dt} \ket{\tilde\psi\pqty{t}} + \bra{\tilde\psi\pqty{t}}\frac{d\ket{\tilde\psi\pqty{t}}}{dt} \\
    &= \mel{\tilde\psi\pqty{t}}{\pqty{iH^\dagger_\mathrm{eff} - iH_\mathrm{eff}}}{\tilde\psi\pqty{t}}\\
    &= -2 \mel{\tilde\psi\pqty{t}}{H_\mathrm{a}}{\tilde\psi\pqty{t}}\\
    &= -2 \mel{\psi\pqty{t}}{H_\mathrm{a}}{\psi\pqty{t}} \braket{\tilde\psi\pqty{t}} \ .
\end{align}
\end{subequations}
The solution to the above equation is
\begin{equation}
    \braket{\tilde\psi\pqty{t}} = e^{-2 \int_0^{t} \mel{\psi\pqty{\tau}}{H_\mathrm{a}}{\psi\pqty{\tau}} d\tau} \braket{\tilde\psi\pqty{0}} \ .
\end{equation}
In the above equation, $e^{-2 \int_0^{t} \mel{\psi\pqty{\tau}}{H_\mathrm{a}}{\psi\pqty{\tau}} d\tau}$ is monotonically decreasing. To see this, we first examine the case of the unravelled Lindblad equation, where $H_a = \frac{1}{2}\sum_k L^\dagger_k L_k$ is positive semi-definite since each term in the summation is quadratic. As a result, $\int_0^{t} \mel{\psi\pqty{\tau}}{H_\mathrm{a}}{\psi\pqty{\tau}} d\tau$ is always a positive value. We then turn to the case of the vectorized Lindblad equation. Because the evolution operator of the original Lindblad equation is a contraction map, i.e.,
\begin{equation}
    \sqrt{\Tr\bqty{\rho^\dagger\pqty{t_2}\rho\pqty{t_2}}} \leq \sqrt{\Tr\bqty{\rho^\dagger\pqty{t_1}\rho\pqty{t_1}}} \ , \quad \mathrm{for} \ t_1 \leq t_2 \ ,
\end{equation}
and the linear isometry preserves the trace norm and $L_2$ norm, we have
\begin{equation}
    \braket{\tilde\psi\pqty{t_2}} \leq \braket{\tilde\psi\pqty{t_1}} \ , \quad \mathrm{for} \ t_1 \leq t_2 \ .
\end{equation}

For a small time interval $dt$
\begin{equation}
    \braket{\tilde\psi\pqty{t+dt}} \approx e^{-2 \mel{\psi\pqty{t}}{H_\mathrm{a}}{\psi\pqty{t}} dt} \braket{\tilde\psi\pqty{t}} \ ,
\end{equation}
thus we can keep track of the norm by measuring $H_\mathrm{a}$ at each time step.

\section{Lower bound of the McLachlan's distance}
\label{app:lower_bound}
In this Appendix, we derive a lower bound of the McLachlan's distance (Eq.~\eqref{app_eq:D_terms}). First, we note from Eq.~\eqref{app_eq:ansatz} that the ansatz state is parameterized by a unitary 
\begin{equation}
    \ket{\phi\pqty{\bth\pqty{t}}} = U\pqty{\bth\pqty{t}} \ket{\psi_\mathrm{R}} \ .
\end{equation}
A reasonable assumption we make is that $d U\pqty{\bth\pqty{t}} / dt$ exists (well-defined and finite). With this assumption, we can define an effective Hamiltonian parameterized by the derivative of the ansatz parameter
\be
\tilde{H}\pqty{\dot{\bth}} = i \dot{U}\pqty{\bth} U^{-1}\pqty{\bth} \ ,
\ee
where we omit the $t$-dependence of each quantity in the above equation. It is straightforward to check the following relation is satisfied
\be
\label{app_eq:eff_Htheta}
\dot{U}\pqty{\bth} = -i \tilde{H}\pqty{\dot{\bth}} U\pqty{\bth} \ .
\ee
Then the derivative of the ansatz state is subject to the following equation of motion
\be
\frac{dW\pqty{\bth}}{dt} = - i\comm{\tilde{H}\pqty{\dot{\bth}}}{W\pqty{\bth}} \ ,
\ee
and the McLachlan's distance becomes
\begin{subequations}
\begin{align}
    \mathcal{D} &= \norm{\frac{dW\pqty{\bth\pqty{t}}}{dt}-\Lag\bqty{W\pqty{\bth\pqty{t}}}}^{2}    \\
    &=\norm{\pqty{\tilde{H}-\Heff}W-W\pqty{\tilde{H}-\Heff^\dagger}}^2 \label{app_eq:H-Heff}\\
    &=\norm{\pqty{\Delta H+i\Ha}W-W\pqty{\Delta H-i\Ha}}^2 \label{app_eq:DH-Ha}\\
    &=2\expval{\Delta H^2} - 2\expval{\Delta H}^2 + 2\expval{\Ha^2} + 2\expval{\Ha}^2 -2i\expval{\comm{\Ha}{\Delta H}} \ ,
\end{align}
\end{subequations}
where we denote $\Delta H\equiv \tilde{H}-\He$ in going from Eq.~\eqref{app_eq:H-Heff} to Eq.~\eqref{app_eq:DH-Ha}.
When $\comm{\Ha}{\Delta H}=0$, a lower bound of $\mathcal{D}$ is given by
\be
\mathcal{D} \ge 2\expval{\Ha^2} + 2\expval{\Ha}^2 \ ,
\ee
where the equality is achieved when $\Delta H=0$, which means the optimal $\dot{\bth}$ makes $\tilde{H}\pqty{\dot{\bth}}$ in Eq.~\eqref{app_eq:eff_Htheta} equal the Hermitian part of the effective Hamiltonian $\Heff$.

A sufficient condition for $\comm{\Ha}{\Delta H}=0$ is
\begin{subequations}
\begin{align}
    \comm{\He}{\Ha} &= 0 \\
    \comm{A_\mu}{\Ha} & = 0 \ , \forall \mu \in \bqty{1, k} \label{app_eq:comm_A}\ ,
\end{align}
\end{subequations}
where Eq.~\eqref{app_eq:comm_A} guarantees $\comm{\tilde{H}\pqty{\dot{\bth}}}{\Ha}=0$ .

\section{Measure observable $O$ in the vectorization method}
\label{app:mO}
In this section we describe the circuit used to measure the observable $O$ for the vectorization method based on Ref.~\cite{Kamakari2022-yr}. The target is to evaluate the quantity $\braket{O^\dagger}{\phi(t)}$ where $\ket{O^\dagger}$ is the vectorized operator $O^\dagger$ from a smaller Hilbert space. The main observation is that the parameterized state $\ket{\phi\pqty{t}}$ can be generated from $\ket{0}$ (all zero state) with one unitary
\begin{equation}
    \ket{\phi\pqty{t}} = \prod_{\mu=1}^{k} e^{-i\theta_\mu\pqty{t} A_\mu} \ket{\psi_\mathrm{R}} = U\pqty{\bth} \ket{\psi_\mathrm{R}} =U\pqty{\bth} U_\mathrm{R}\ket{0} \ ,
\end{equation}
where $U_\mathrm{R}$ prepares the reference state.
If we can construct another unitary that prepares $\ket{O^\dagger}$, i.e. $\ket{O^\dagger} = V \ket{0}$, then $\braket{O^\dagger}{\phi(t)}=\mel{0}{V^\dagger U\pqty{\bth}U_\mathrm{R}}{0}$ can be measured either directly or indirectly (Hadamard test). Assume the operator $O$ has dimension $2^N\times 2^N$, an explicit form of $\ket{O^\dagger}$ is given by
\begin{equation}
    \label{app:vec_O}
    \ket{O^\dagger}= \sum_{x_1 y_1}\frac{O^*_{x_1 y_1}}{\sqrt{\Tr\pqty{O^\dagger O}}}\ket{x_1y_1} \ ,
\end{equation}
where $x_1$ and $y_1$ are bit strings for $0$ to $2^N-1$ and $O_{x_1 y_1}$ is the $x_1$, $y_1$ element of the operator. $V$ can be synthesized using quantum state preparation algorithm~\cite{Sun2023-mc,szasz2023numerical}. Because $O$ is usually a local observable, Eq.~\eqref{app:vec_O} can be further simplified to avoid any exponential cost~\cite{Kamakari2022-yr}.  In the examples considered in the main text, the reference state is chosen as $\ket{\psi_\mathrm{R}}=\ket{+}^{\otimes 2N}$, as a result, $U_\mathrm{R}$ can be implemented using Hadamard gates $H^{\otimes 2N}$.

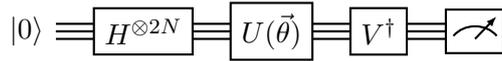
\begin{figure}[htb]
    \centering
    \begin{quantikz}
        \lstick{$\ket{0}$} &  \gate{H^{\otimes 2N}}\qwbundle[
        alternate]{} & \gate{U(\vec{\theta})} \qwbundle[
        alternate]{}
        & \gate{V^\dagger}\qwbundle[
        alternate]{} & \meter{}\qwbundle[
        alternate]{}
    \end{quantikz}
    \caption{Direct measurement scheme to evaluate $\braket{O^\dagger}{\phi(t)}$. The final result can be obtained by taking the square root of the all-$0$ string frequency.}
    \label{fig:vectorized_measure}
\end{figure}

In our experiments, we used the Berkeley Quantum Synthesis Toolkit (BQSKit)~\cite{Younis2021-qs,szasz2023numerical} to synthesize the unitary $V$ and used the direct measurement protocol shown in Fig.~\ref{fig:vectorized_measure} to evaluate $\braket{O^\dagger}{\phi(t)}$. The Hadamard test is not currently practical on hardware because it requires implementing $V^\dagger U\pqty{\vec{\theta}}H^{\otimes 2N}$ as a control unitary. We can obtain the final result by taking the square root of the all-$0$ string frequency since $\braket{O^\dagger}{\phi(t)} = \ev{V^\dagger U\pqty{\vec{\theta}} H^{\otimes 2N}}{0}$. When taking the square root, a sign needs to be assigned; however, we can always shift an observable by the identity matrix such that all its eigenvalues have the same sign.
\section{Implementing $L_-$ and $L_+$ jump operators}
\label{app:jumps}
\begin{figure}[htb]
    \centering
    \subfloat[]{
    \begin{quantikz}
    \lstick{$\ket{\phi}$} & \meter{$I/X$} \arrow[r] & \rstick{$\ket{0}$}
    \end{quantikz}
    }%
    \qquad
    \subfloat[]{%
   \begin{quantikz}
    \lstick{$\ket{0}$}& \targ{} & \meter{} \arrow[r]& \rstick{$\ket{0}$} \\
    \lstick{$\ket{\phi}$}& \ctrl{-1}  &\qw &\qw
    \end{quantikz}
    }%
    \caption{Quantum circuit for implementing an eraser channel $\mathcal{E}\pqty{\rho}=\dyad{0}$ using (a) mid-circuit measurement; (b) block encoding. The measurement symbol in (a) means a conditional gate based on the measurement result. We only apply $X$ gate when the result is $1$. The arrow pointing to $\ket{0}$ in (b) represents the post-selection. We only keep the runs where $0$ is returned by the measurement.} 
    \label{fig:q_eraser}
\end{figure}
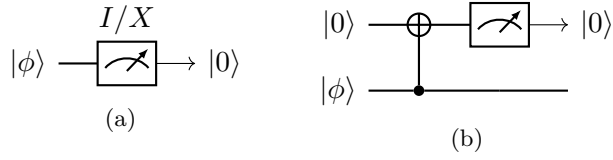

In this appendix, we describe two methods for applying the $L_+$ and $L_-$ jump operators on a quantum computer. Without loss of generality, we present only the derivation for $L_+\equiv (X+iY)/2$, and the procedure is easily extended to $L_-\equiv (X-iY)/2$. The key observation is that the $L_+$ jump only occurs when its probability $p_+\propto \ev{L_-L_+}{\phi}$ is greater than zero, meaning $\ket{\phi}\neq \ket{0}$. Under this condition, the action of the jump operator $L_+\ket{\phi}/\sqrt{\ev{L_-L_+}{\phi}}$ is equivalent to a quantum eraser channel $\mathcal{E}\pqty{\rho}=\dyad{0}$. There are two ways of implementing such a channel on a quantum computer: mid-circuit measurement or block encoding (unitary dilation)~\cite{Low2017-ye,camps2023explicit,Del_Re2020-ul}. The former requires us to perform a mid-circuit measurement on the qubit and apply a $X$ gate conditioned on the result $1$. The latter requires an additional ancilla qubit prepared in $\ket{0}$ state. Then we apply a CNOT gate with the ancilla as the target qubit, measure the ancilla and post select the runs with measurement result $0$. The circuits for both of these methods are shown in Fig.~\ref{fig:q_eraser}.
\section{Growth of ansatz size}
\label{app:growth_ansatz}
\begin{figure}[htb]
    \centering
    \subfloat[]{
    \label{fig:traj_N_G}
    \includegraphics[width=0.47\linewidth]{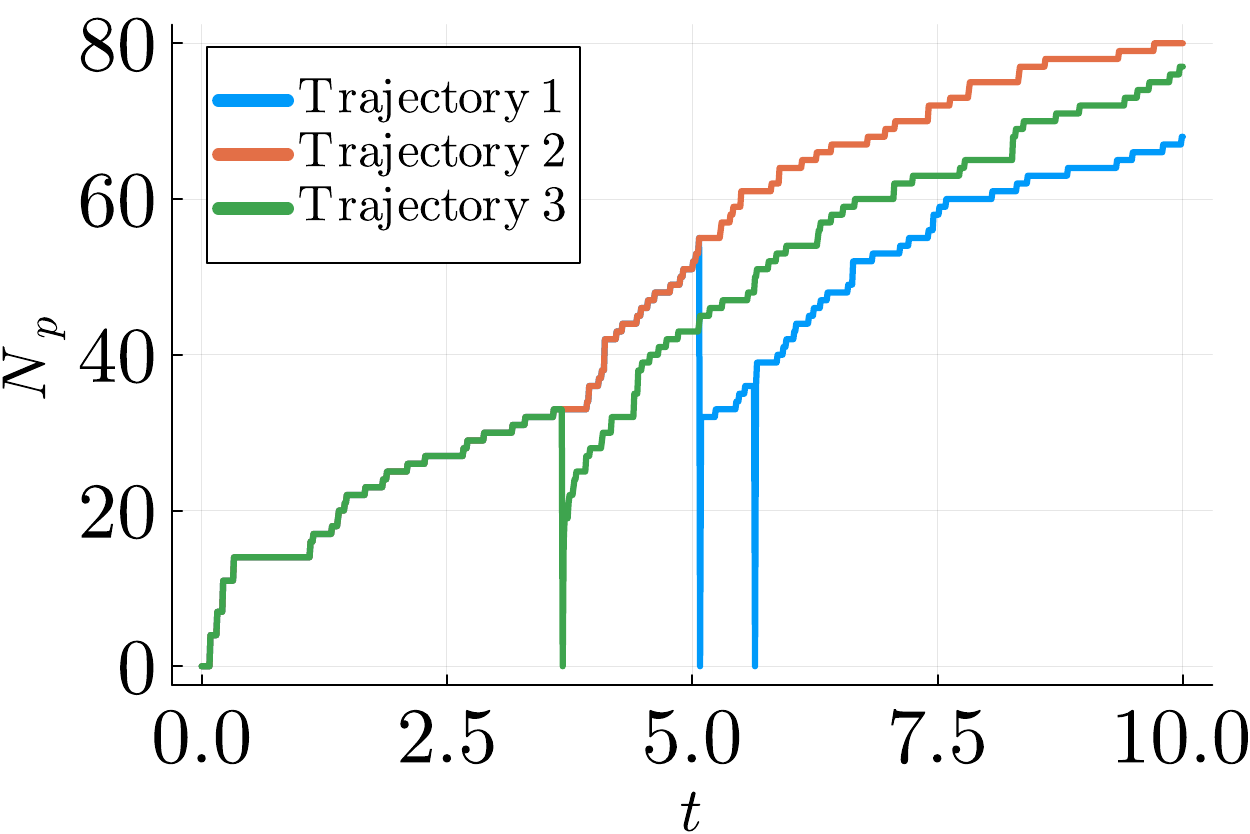}
    }
    \hfill
    \subfloat[]{%
    \label{fig:vec_N_G}
    \includegraphics[width=0.47\linewidth]{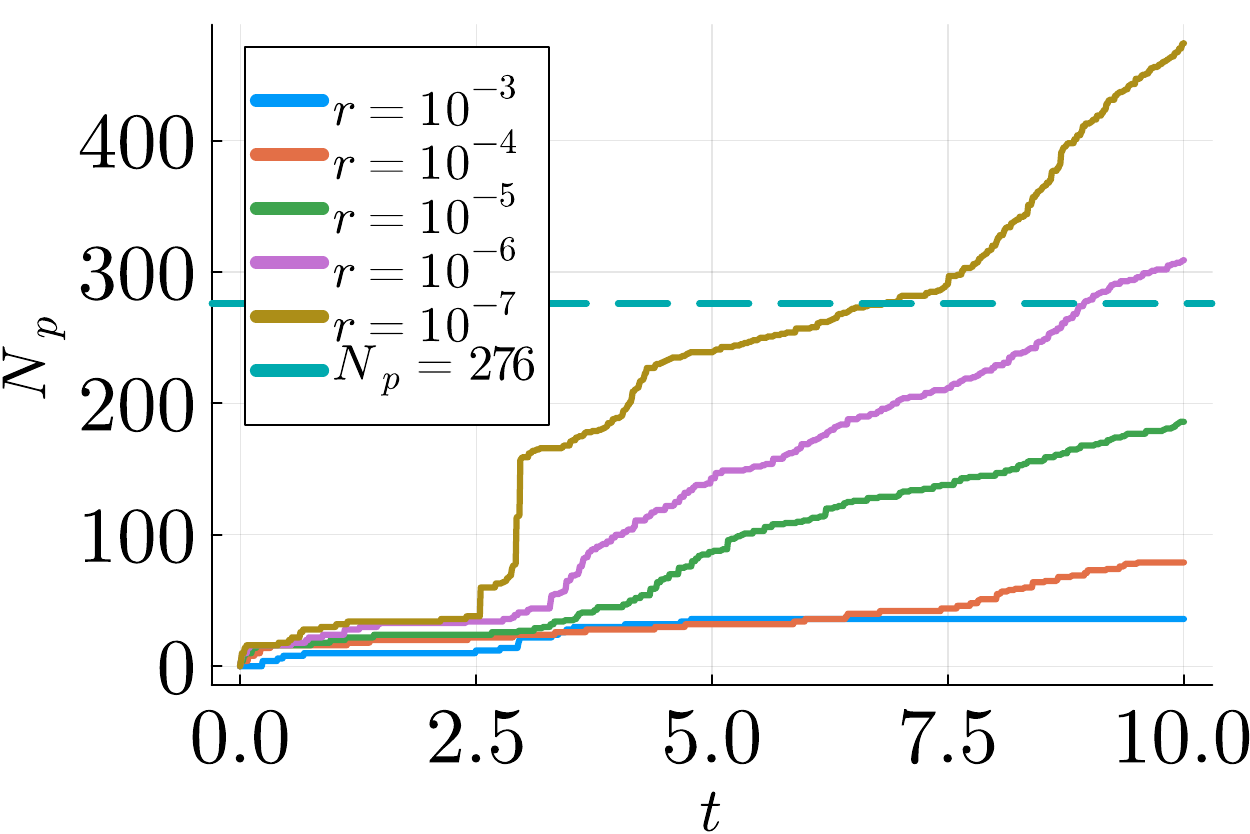}
    }%
    \caption{\textbf{Number of ansatz parameters required for different evolution times in both the trajectory and vectorization methods.} In (a), the numbers of parameters are plotted for three different trajectories, with vertical lines indicating jump events where the ans\"atze are reset. In (b), the numbers parameters for the vectorization method is shown for different error thresholds, with the turquoise dashed line indicating the total number of distinct operators in the operator pool.}
    \label{fig:n4_param_grow}
\end{figure}

In this appendix, we demonstrate how the sizes of ans\"atze grow as we adaptively add operators to them. The results are obtained from the examples with $N=8$ showcased in Fig.~\ref{fig:nscaling_t1} and $N=4$ in Fig.~\ref{fig:nscaling_v1} in the main text. In Fig.\ref{fig:traj_N_G}, we present examples of three different trajectories where $0$, $1$, and $2$ jump events occurred during the evolution. As we can see, the adaptive procedure quickly refills the ans\"atze after they are reset by the jumps. Consequently, the accumulated ansatz size becomes larger when more jumps occur. In Fig.\ref{fig:vec_N_G}, we showcase examples using the vectorization method with different error thresholds. We observe larger ansatz sizes when lower threshold values are used.

\section{The vectorization method with different operator pools and error threshold}
\label{app:vec_figs}
In this appendix, we display the average energy $\expval{H(t)}$ versus evolution time $t$ for both the dephasing and amplitude damping models obtained using the vectorized UAVQDS with different choices of operator pools and error thresholds. The purpose is to provide visualization of how the error behaves with respect to various operator pools and error thresholds.
\begin{figure}[h]
    \centering
    \subfloat[]{
    \includegraphics[width=0.45\linewidth]{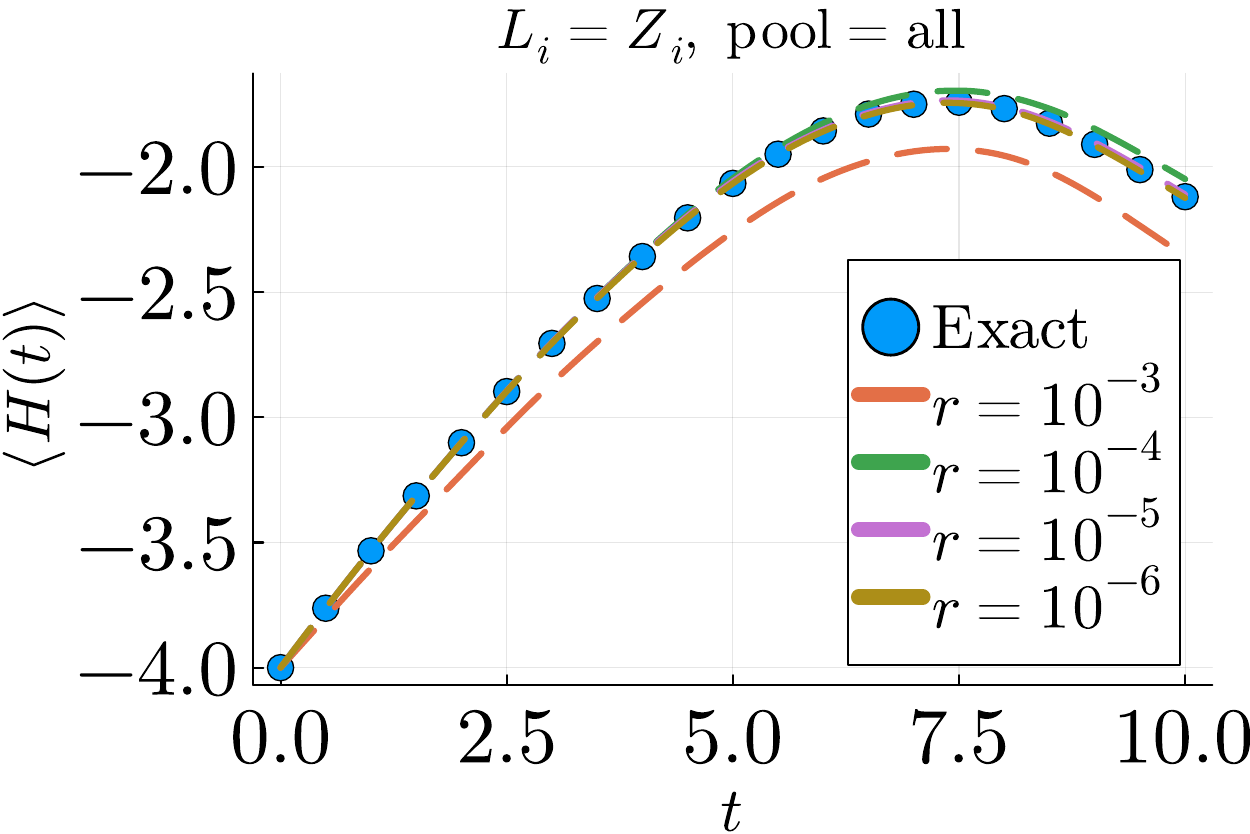}
    }
    \hfill
    \subfloat[]{%
    \includegraphics[width=0.45\linewidth]{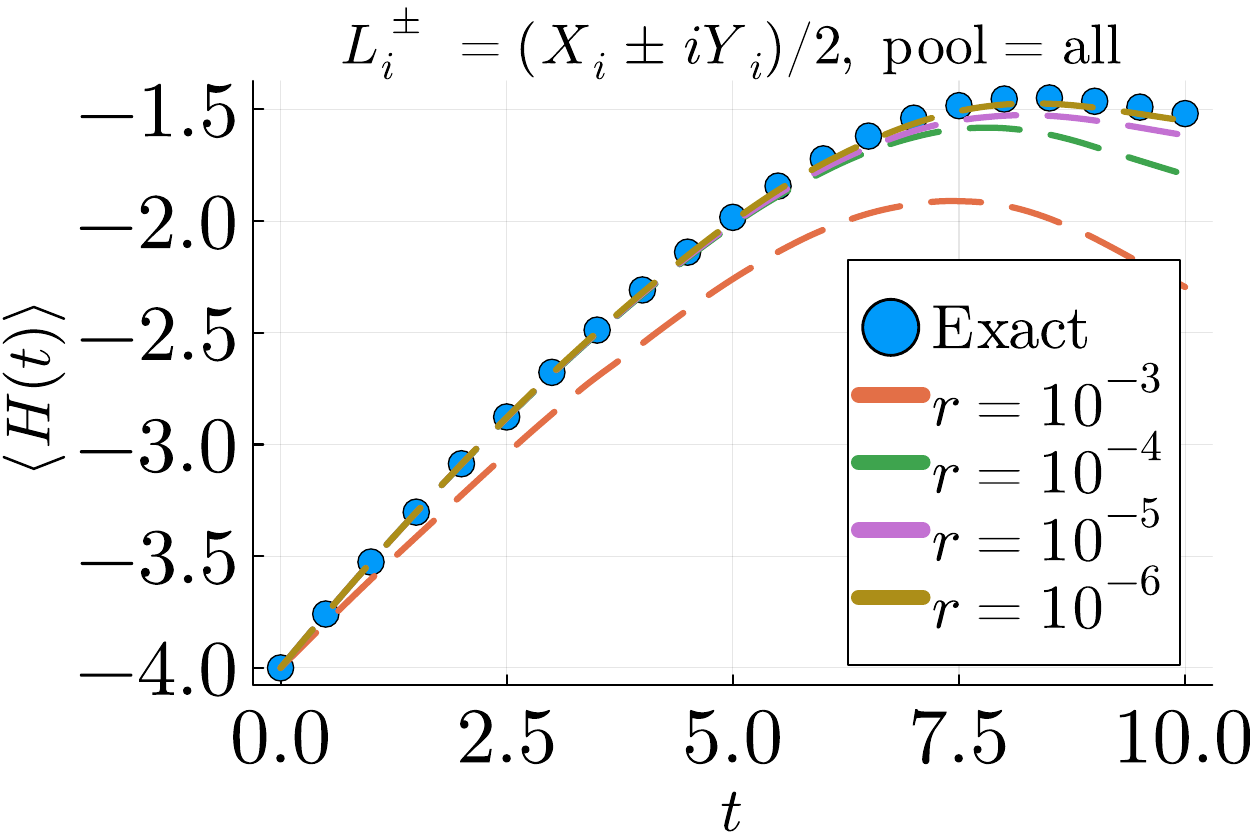}
    }%
    \\
    \subfloat[]{
    \includegraphics[width=0.45\linewidth]{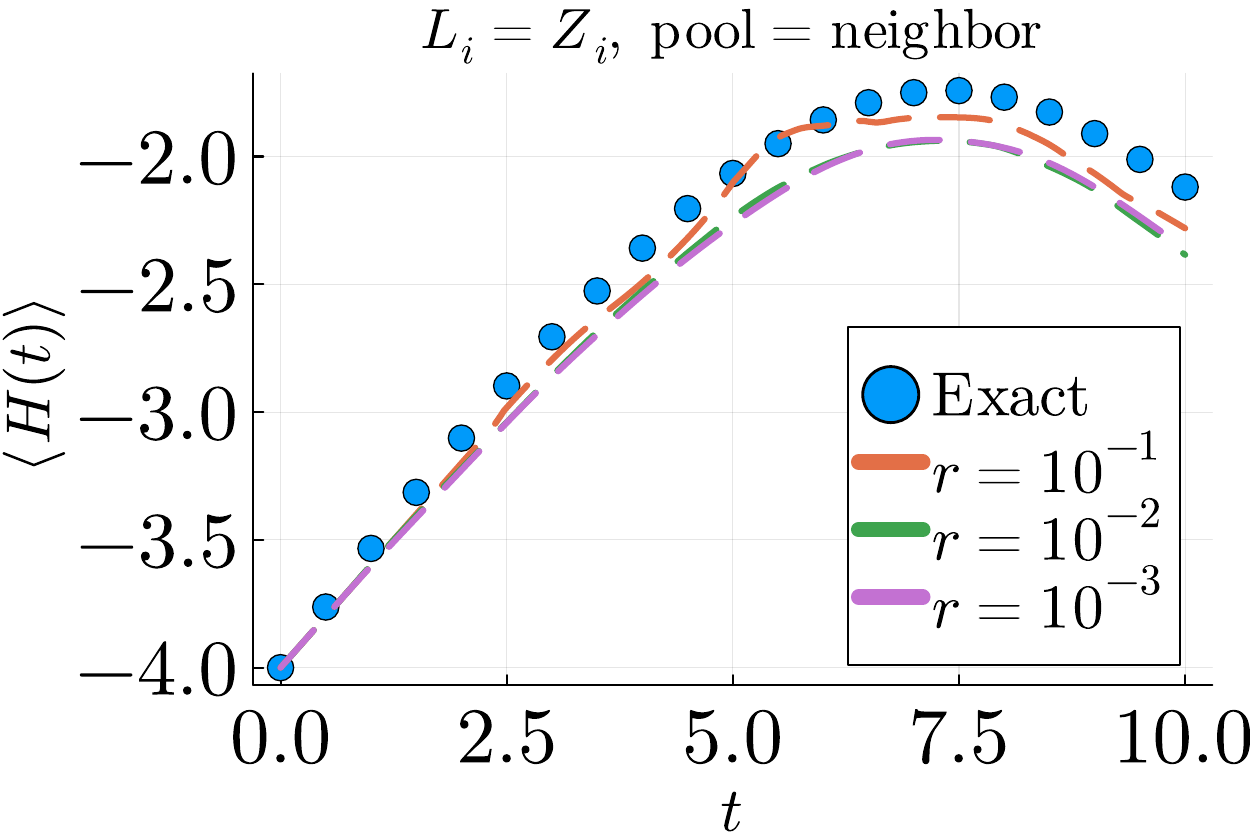}
    }
    \hfill
    \subfloat[]{
    \includegraphics[width=0.45\linewidth]{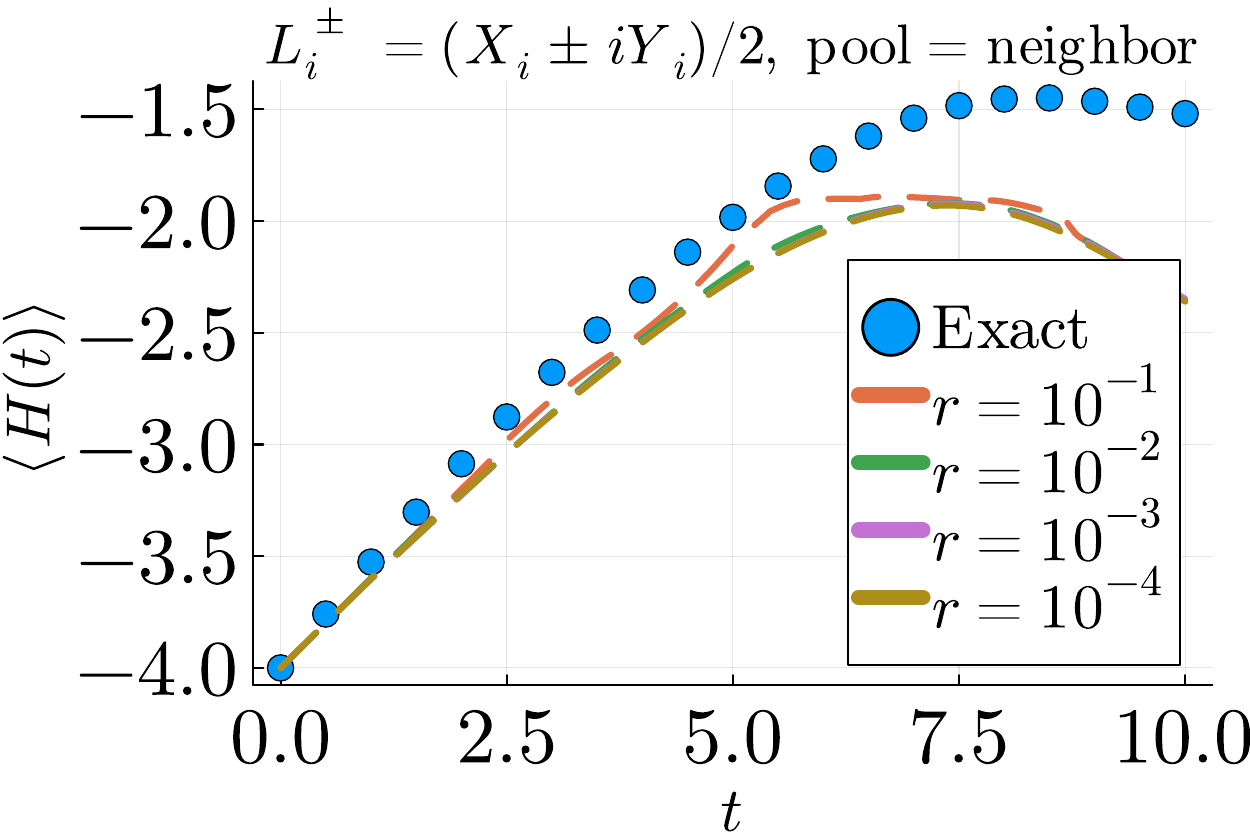}
    }%
    \caption{\textbf{The energy evolution of the vectorization UAVQDS for different Lindblad models, error thresholds and operator pools.} Subplots (a) and (b) show the results obtained using a all-to-all 2-qubit Pauli pool ($\mathcal{P}_3$ in the main text) for the dephasing and amplitude damping models, respectively. Subplots (c) and (d) show the results obtained using a neighboring 2-qubit Pauli pool ($\mathcal{P}_2$ in the main text) for the dephasing and amplitude damping models, respectively.}
    \label{fig:n4_err_demo}
\end{figure}

\section{ Resolution enhancement technique}
Our method for resolution enhancement relies on the assumption that the bitstring data produced during a noisy experiment somewhat conforms to the expected probability distribution of bitstrings. In simpler terms, even though noise and measurement errors cause the histogram of bitstrings to become less distinct, the underlying pattern of the histogram remains intact. To align the bitstring distribution with the ideal one, we apply resolution enhancement techniques commonly employed in image processing \cite{Ohaver2022}. The corrected bistring distribution data is then used to compute the expectation values.  While the details of the method have been already described in \cite{gomes2023computing},
for the sake of completeness we briefly discuss the method here as well as provide the specific steps we have undertaken for our current work.

In our work, we measure a single Pauli term  using Hadamard test circuit and later add multiple expectation values to find the energy expectation value $\expval{H(t)}$. Hadamard test requires one ancilla qubit and the expectation value can be obtained by just measuring the ancilla qubit. We, however, measure all physical qubits (including the ancilla) and then trace over the rest to obtain the state of the ancilla. Such additional measurements also help getting rid of certain unphysical states that arise in quantum experiments, e.g, keeping track of fermion number conservation \cite{steckmann2023}. In our work such screening were not necessary. Writing explicitly, if the measured wavefunction of the full system is given by $\ket{\psi} = \sum_j \alpha_j\ket{b_n ..b_1}\ket{0}_a + \beta_j\ket{b_n..b_1}\ket{1}_a$, where $b_j$s are the measured state of of the qubits, after applying our RE method, we will obtain a corrected wavefunction  $\ket{\psi_c} = \sum_j \alpha_j^{(c)}\ket{b_n ..b_1}\ket{0}_a + \beta_j^{(c)}\ket{b_n..b_1}\ket{1}_a$.

Calling $y_j = \abs{\alpha_j}^2$ as the frequency of the noisy data of $j$-th bitstring, resolution enhanced frequency is obtained using $r_j = y_j - k_2 y_j'' $, were $r_j= ( \abs{\alpha_j^{(c)}}^2)$ is the reformed frequency and $y_j''$ is the second derivative of the noisy data w.r.t the decimal representation of the bitstrings. The parameter $k_2$ can be modified to tune the resolution of the final data. 
\begin{figure*}[t]
    %\centering
    \includegraphics[width=\linewidth]{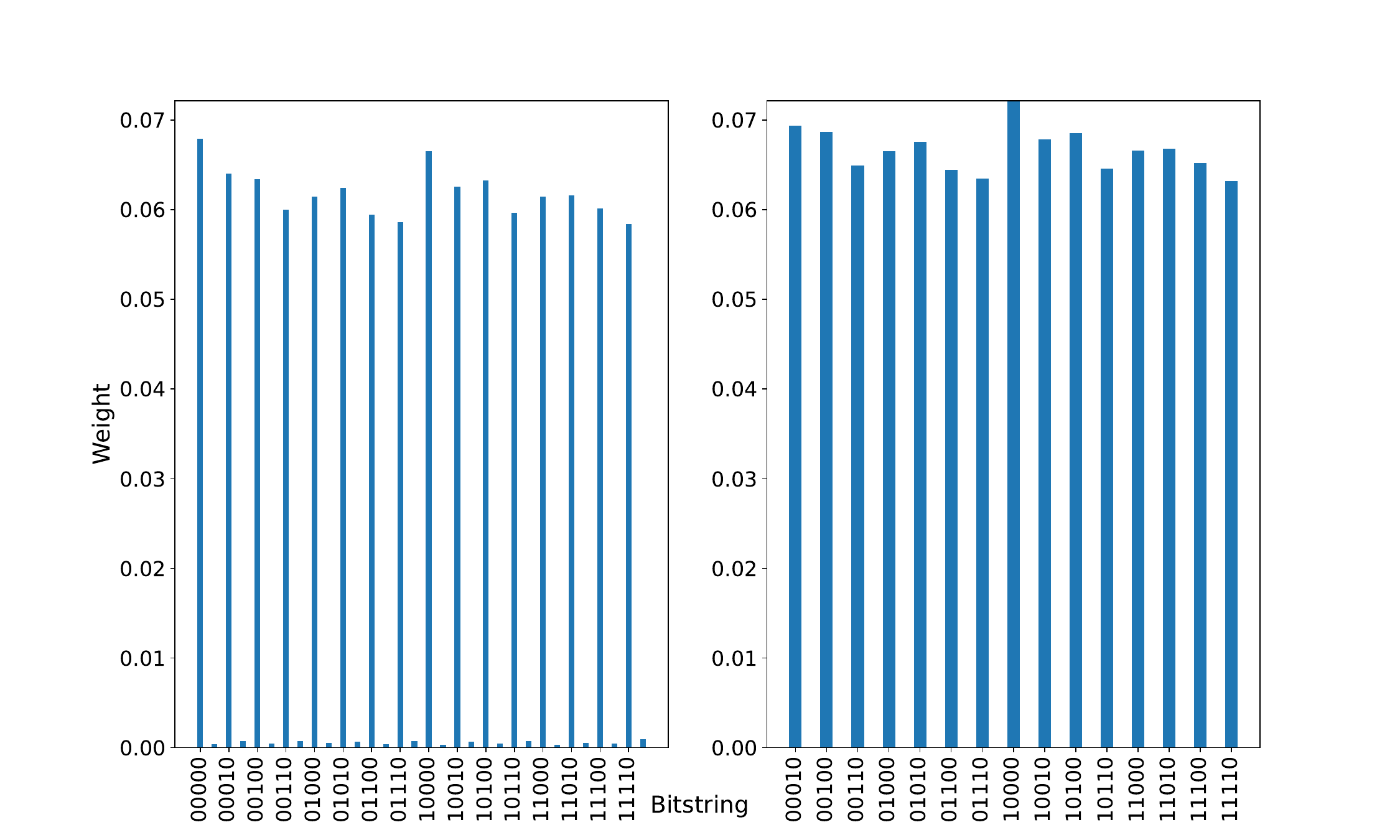}
    \caption{\textbf{Effects of resolution enhancement on bitstring distribution.} Histogram of bitstrings of noisy simulation (a) before and (b) after applying resolution enhancement }   
    %source file: dchp: /home/niladri/rcut_a_N4_hubbard/final_green
    %source data: cori: /global/homes/n/niladri9/hardware_gf/rcut_a_N2_hubbard/final_green
    \label{fig:res_en}
\end{figure*}
The weighting factor $k_2$ can be chosen based on what gives the best trade-off between resolution enhancement, signal-to-noise degradation, and baseline flatness. The optimum choice depends upon the width, shape, and digitization interval of the the bitstrings. After obtaining $r_j$ we identify the bitstrings nearest to the peaks from the resolution enhanced data. We then switch back to the binary representation of the bitstrings and replace $y_j$s by the $r_j$s. 

In order to avoid the ambiguity of the optimal value of $k_2$, we iterate over several values of $k_2$ and calculate the probability $p_0$ of the ancilla qubit for each of them. We continue iterating until  $p_0$ converges with a certain threshold $\epsilon$. In other words, if $p_0(k_2^{(j)})$ is the probability at the $j-$th iteration, we first calculate the average of  $p_0(k_2^{(j)})$ over the previous $j$ values of $k_2$. Thus we may define,
\be  
\mathbf{\hat{p}_0}(k_2^{(j)}) = \frac{1}{j}\sum_{l=1}^{j}p_0(k_2^{(l)})
\ee  
we stop the loop if $\abs{\mathbf{\hat{p}_0}(k_2^{(j+1)}) - \mathbf{\hat{p}_0}(k_2^{(j)})} < \epsilon$. For our calculation we chose $\epsilon\sim 10^{-4}$ and varied $k_2$ in the range $[0,8]$ in steps of $0.1$. To understand the effect of the method, we show the results in Fig.~\ref{fig:res_en},  where the original noisy data is shown in the left panel and right panel shows after the resolution enhancement is applied. The plot shows the bitstring data for the $N=4$ calculation and,  we have chosen the eleventh trajectory of our experiment as an example.  The histogram in the right panel has the corrected peaks. 

\end{document}